\documentclass[aip,jcp,reprint]{revtex4-1}

\usepackage{graphicx}
\usepackage{bm}

\newcommand{\ve}[1][K]{\mathbf{#1}}

\usepackage{amsmath}
\usepackage{amssymb}
\usepackage{latexsym}
\usepackage{multirow}

\begin{document}

\title{Reactive conformations and non-Markovian cyclization kinetics of a Rouse polymer}

\author{T. Gu\'erin}
\affiliation{Laboratoire de Physique Th\'eorique de la Mati\`ere Condens\'ee, CNRS/UPMC, 
 4 Place Jussieu, 75005 Paris, France.}

\author{O. B\'enichou}
\affiliation{Laboratoire de Physique Th\'eorique de la Mati\`ere Condens\'ee, CNRS/UPMC, 
 4 Place Jussieu, 75005 Paris, France.}

\author{R. Voituriez}
\affiliation{Laboratoire de Physique Th\'eorique de la Mati\`ere Condens\'ee, CNRS/UPMC, 
 4 Place Jussieu, 75005 Paris, France.}

\begin{abstract}
We investigate theoretically the physics of diffusion-limited intramolecular polymer reactions. 
The present work completes and goes beyond a previous study [Nat. Chem. \textbf{4}, 268 (2012)] that showed that the distribution of the polymer conformations at the very instant of reaction plays a key role in the cyclization kinetics, and takes explicitly into account  the non-Markovian nature of the reactant motion. Here, we present in detail this non-Markovian theory, and compare it explicitly with existing Markovian theories and with numerical stochastic simulations. A large focus is made on the description of the non-equilibrium reactive conformations, with both numerical and analytical tools. We show that the reactive conformations are elongated and are characterized by a spectrum with a slowly decreasing tail, implying that the monomers that neighbor the reactive monomers are significantly shifted at the instant of reaction. 
We complete the study by deriving explicit formulas for the reaction rates in the Markovian Wilemski-Fixman theory  when the reactants are located in arbitrary positions in the chain.
We also give a simple scaling argument to understand the existence of two regimes in the reaction time, that come from two possible behaviors of monomer motion which can be either diffusive or subdiffusive. 
\end{abstract}


\bibliographystyle{apsrev}


\maketitle

\section{Introduction}

Determining how fast two reactants attached to a polymer come into contact is an old problem of statistical mechanics \cite{WILEMSKI1974b,WILEMSKI1974a,Pastor1996,Szabo1980,Friedman1989}. 
When a reactant molecule is attached to a polymer, its interaction with the whole polymer chain results in a complex motion that can be subdiffusive \cite{KhokhlovBook,DoiEdwardsBook} and  leads to non-trivial reaction kinetics \cite{DEGENNES1982,Nechaev:2000fk,OSHANIN1994}. 
The cyclization reaction between the two end monomers is an important example of intramolecular reaction and has been extensively studied, both theoretically  \cite{WILEMSKI1974a,WILEMSKI1974b,Szabo1980,Friedman1989,FRIEDMAN1993b,FRIEDMAN1993,DEGENNES1982,Toan2008} and experimentally, for example in the context of hairpin formation in nucleic acids \cite{Bonnet1998,Wallace2001,Wang2004,Uzawa2009} or the folding of polypetide chains \cite{Lapidus2000,Moglich2006,Buscaglia2006}. Indeed, the cyclization of a polypeptide chain can be seen as an elementary step of  the folding pathway \cite{Lapidus2000}. In the context of nucleic acids, the formation of loops and hairpins in DNA is a key process in the regulation of gene expression\cite{Allemand2006}, while the cyclization of molecular beacons can be used as a tool for the recognition of nucleic acid sequences\cite{Bonnet1998,Errami2007}. More generally, the cyclization time is of interest for example because it controls in part the formation of rings in the polymerization reactions\cite{Flory1971}.

In this paper, we study the diffusion controlled reaction between two reactive groups that are attached to a single polymer chain. The theoretical description of polymer reaction kinetics is made complicated by the structural dynamics of the chain:
the position of a single monomer cannot be described as a Markov process, because it results from the interactions between all the monomers of the chain. Here we focus on the non-Markovian effects and we consider the simple case where the polymer is modeled by a Rouse chain of beads and springs where both hydrodynamic and excluded volume interactions are neglected. Despite its simplicity, this model catches some important aspects of polymer dynamics\cite{KhokhlovBook,DoiEdwardsBook}, and the calculation of the mean cyclization time is not trivial due to the presence of the non-Markovian effects that we aim at  describing. Classical approaches on polymer reaction kinetics rely on ``Markovian approximations'', such as the harmonic spring approximation\cite{Szabo1980,SUNAGAWA1975,Szabo1980} (where the polymer is approximated by a single spring and the problem is therefore Markovian), or the more refined closure approximation in the Wilemski-Fixman theory\cite{WILEMSKI1974b,WILEMSKI1974a}, which is a local equilibrium assumption for the whole polymer. Because memory effects appear at the largest relaxation time scale of the polymer, which can be of the same order of magnitude as the average cyclization time, Markovian approximations have inevitably a restricted range of validity\cite{Pastor1996}. 
Alternative theoretical approaches have been proposed, such as the use of the renormalization group theory \cite{Friedman1989,FRIEDMAN1993b,FRIEDMAN1993} which provides for long chains perturbative results for small values of $4-d$, with $d$ the spatial dimension. More recent approaches include a refinement of the Wilemski-Fixman theory that considers the correlations between the initial and the final states\cite{Sokolov2003,Campos2012}, an exact formal iterative solution to the cyclization problem in one dimension\cite{Likthman2006}, or a derivation of the cyclization time in the limit of a very small reactive radius from first principles\cite{Amitai2012}. 
 
\begin{figure}[h!]
\includegraphics[width=7cm,clip]{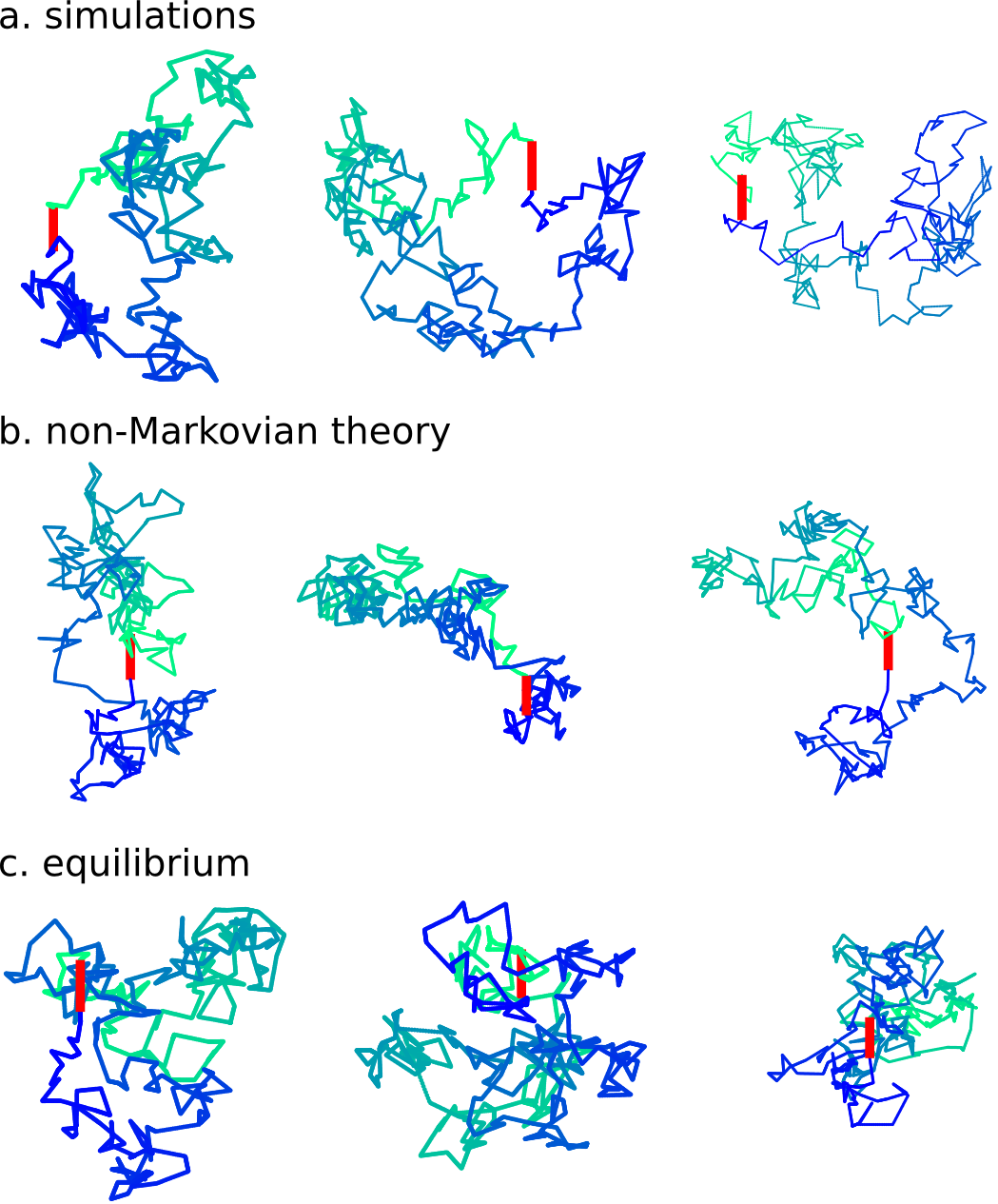}
\caption{Examples of polymer conformations at the instant of first contact between the two ends of a Rouse chain. The conformations in (a) are obtained from Brownian dynamics simulations, while the conformations in (b) are the results of the non-Markovian theory that is described in this paper. In (c), we represented equilibrium looping conformations. The fact that the reactive conformations (a,b) tend to be more elongated than equilibrium looping conformations (c) leads to a faster reaction kinetics than predicted by the Wilemski-Fixman Markovian theory. 
The red thick vertical line represents the end-to-end vector $\ve[R]$. The number of monomers is $N=300$ monomers and the capture radius is $a=4.33$. Precise statistics of reactive conformations for the same parameters are presented in Figs. \ref{FigHistogrammesModes}, \ref{FigSIMADiag}, \ref{SpectrumPolymerCyclization} and \ref{ShapePolymerCyclization}.
\label{ExamplesConfigurationsCyclization}}
\end{figure}  
 
In a recent work\cite{Guerin2012}, we proposed another approach of the problem, in which non-Markovian effects are explicitly taken into account by determining the statistics of the  non-equilibrium polymer conformations at the very instant of the reaction. This approach is referred to below as the non-Markovian theory. 
Examples of reactive conformations obtained by this approach are shown on Fig. \ref{ExamplesConfigurationsCyclization}, where one observes that a polymer tends to be more elongated at the instant of cyclization than in an equilibrium looping conformation. 
This elongated shape leads to reaction times that are faster than predicted by the Wilemski-Fixman Markovian approach. The main goal of the present paper is to complete this non-Markovian theory of polymer intramolecular reaction kinetics, and to present new results that are related to it. In particular, we provide an analytical description of the average reactive conformation of the polymer, and we describe the non-Markovian theory in detail. We also review the existing approaches of cyclization in the context of our non-Markovian approach and of stochastic simulations. Establishing the validity range of the Markovian  theories is important because they are frequently invoked in the analysis of experiments on hairpin formation or the folding of polypeptide chains \cite{Lapidus2000,Wallace2001,Buscaglia2006,Moglich2006}. Finally, we also give explicit formulas that describe the effect of the position of the reactants in the chain for the reaction kinetics, and we show that the $j$-factor, defined as the equilibrium contact probability density between the reactants\cite{Allemand2006,Rippe2001} appears naturally in the expression of the reaction time but is not the only determinant of diffusion controlled reaction kinetics. We describe in the framework of the Wilemski-Fixman approximation how the reaction time is slowed when the reactive monomers belong to the interior of the chain. 
The present work completes another paper dealing with intermolecular reactions\cite{guerin2012b}. 

The outline of this paper is at follows. In section \ref{SectionDefinitionRouseChain}, we introduce the formalism of the Rouse chain, and we give a simple argument that enables us to derive a scaling relation for the reaction time. Then, in section \ref{SectionNonMarkovianTheory}, we present in detail a non-Markovian theory of intramolecular polymer reaction kinetics. 
In section \ref{SectionComparisonTheories}, we discuss the validity of the hypotheses of existing Markovian theories in the context of our non-Markovian theory. 
In section \ref{SectionComparisonTheoriesWithSimus}, we determine the accuracy of the non-Markovian theory and of existing theories by confronting them to the results of stochastic simulations. 
In section \ref{SectionReactiveShapeAndPositionMonomers}, we study in details the reactive shape of the polymer both with our non-Markovian theory and with simulations. Finally, in section \ref{SectionEffectPositionReactiveMonomers}, we investigate the role of the position of the reactive monomers. 
This work is completed by appendices, where we remind some useful properties of Gaussian processes and we give details of calculations.   

\section{Definition of the Rouse chain and simple scaling expression for the reaction time}
\label{SectionDefinitionRouseChain}

We consider the Rouse model of a polymer chain evolving in a three dimensional (3D) space. The chain is formed by $N$ monomers located at positions $\ve[r]_i \ (1\le i\le N)$, in which quantities in bold denote vectors in 3D. The monomers are connected by linear springs of stiffness $k$. Each monomer is submitted to a friction force (with $\zeta$ the drag coefficient) and diffuses with a diffusion coefficient $D=k_BT/\zeta$, where $T$ is the temperature. The length $l_0=\sqrt{k_B T/k}$ is the typical length of a bond, while $\tau_0=\zeta/k$ is the typical relaxation time of an individual bond. In the present paper, the microscopic length $l_0$ and the microscopic time $\tau_0$ are set to $1$, which fixes the units of length and time. We are interested in the kinetics of a reaction that occurs between two reactive monomers of the same chain, that have the indexes $p$ and $q$. 
We define the vector $\ve[R]$ that joins the positions of the reactive monomers:
\begin{align}
	\ve[R]\equiv \ve[r]_q-\ve[r]_p.\label{DefinitionR}
\end{align}
In the diffusion controlled regime, the kinetics of the reaction between the monomers $p$ and $q$ is quantified by the mean time $T$ it takes for the 
distance $\|\ve[R]\|$ to reach a value that is smaller than a typical reaction radius $a$. Of course, the reaction time depends on the choice of initial conditions of the chain. We call $T$ the reaction time. Here, we will focus on the case where the polymer is initially at thermal equilibrium, conditional to the fact that the vector $\ve[R]$ is outside the reactive region ($\|\ve[R]\|\ge a$). 
Note that, although the Rouse model is highly simplified because it neglects both hydrodynamic interactions and excluded volume interactions, it catches some important aspects of polymer dynamics  \cite{KhokhlovBook,DoiEdwardsBook}, and the calculation of intramolecular reaction times is non-trivial\cite{Pastor1996}. The fact of studying the Rouse model enables us to focus on the non-Markovian effects, but our theory could be generalized to more complex polymer models. 

Before exposing the theoretical determination of the reaction time, we remind some characteristics of the dynamics of $\ve[R](t)$ at different time scales that will be useful in the non-Markovian theory.
The dynamics of $\ve[R](t)$ can be characterized by considering the evolution of the monomer positions in terms of independent Rouse modes. The definition of the Rouse modes $\ve[a]_i$ follows from the diagonalization of the connectivity matrix\cite{DoiEdwardsBook,guerin2012b}, and these modes are related to the positions by a linear relation:
\begin{equation}
	\ve[r]_i= \sum_{j=1}^{N} Q_{ij}\ve[a]_j, \label{DefinitionModes}
\end{equation}
where the transfer matrix $Q_{ij}$ reads:
\begin{align}
Q_{ij}=\sqrt{\frac{2-\delta_{1j}}{N}}\cos\left[(i-1/2)(j-1)\pi/N\right].\label{DefinitionMatrixQ}
\end{align}
A polymer configuration can therefore be described either as a set of positions $\vert \ve[r]\rangle=(\ve[r]_1,...\ve[r]_N)$, or as a set of the values of the Rouse modes $\vert \ve[a]\rangle=(\ve[a]_1,...,\ve[a]_N)$. In this paper, we adopt the convention that $\vert u\rangle$ represents a $N$-components column vector. 
The modes $\ve[a]_i$ evolve according to the Fokker-Planck equation:
\begin{equation}
	\frac{\partial P}{\partial t}=\sum_{i=1}^{N} \frac{\partial}{\partial \ve[a]_i}\left(\lambda_i   \ \ve[a]_iP+\frac{\partial}{\partial \ve[a]_i}P\right),\label{FKPRouseModes}
\end{equation}
where the relaxation times $1/\lambda_i$ are the inverses of the eigenvalues of the connectivity matrix, which are
given for $1\le i\le N$ by the relation:
\begin{align}
\lambda_i=2\{1-\cos[(i-1)\pi/N]\}. \label{DefinitionEigenvalues}
\end{align}
The first eigenvalue is $\lambda_1=0$: the first mode $\ve[a]_1$ is proportional to the polymer center-of-mass position, which has a diffusive motion. We call $\tau_R=\lambda_2^{-1}\simeq N^2/\pi^2$ the largest relaxation time of the polymer. The second eigenmode $\ve[a]_2$ is associated to this time scale, and describes the shape of the polymer at the typical length scale $\sqrt{N}$. The other modes $\ve[a]_i$ describe the shape of the polymer at intermediate length scales between $\sqrt{N}$ (the size of the polymer) and $1$ (the size of a single bond). When $N\rightarrow\infty$, the Rouse modes are simply proportional to the Fourier coefficients of the function $\ve[r](s)$, where $s$ is the curvilinear coordinate of a monomer in the chain.

From (\ref{DefinitionR}),(\ref{DefinitionModes}), we observe that we can express the vector $\ve[R]$ as a linear combination of the Rouse modes, which leads to the definition of coefficients $b_i$ by the relation:
\begin{align}
	\ve[R]=\sum_{i=1}^{N} b_i \ve[a]_i \hspace{0.5cm}  \ ; \hspace{0.5cm} \ b_i=Q_{qi}-Q_{pi}. \label{ExpressionCoefficient}
\end{align}
Importantly, the first coefficient vanishes ($b_1=0)$, which means that the dynamics of the vector $\ve[R]$ is independent on the position of the polymer center-of-mass.
At long times, the process $\ve[R](t)$ reaches a stationary state. We call $L$ the equilibrium distance between the reactive monomers, so that $L^2$ is the variance of each spatial coordinate of $\ve[R]$ at equilibrium (the variance of $\ve[R]^2$ at equilibrium is $3L^2$). From (\ref{FKPRouseModes}), we note that  the variance of each coordinate of $\ve[a]_i$ at large times is $1/\lambda_i$, and we deduce the value of the equilibrium length $L$:
\begin{align}
L^2=\sum_{i=2}^N \frac{b_i^2}{\lambda_i}= \vert p-q\vert, \label{ValueOfEquilibriumLength}
\end{align}
where the last equality follows from the use of the explicit expression (\ref{ExpressionCoefficient}) for the coefficients $b_i$. Equation (\ref{ValueOfEquilibriumLength}) simply states that the equilibrium distance between monomers of indexes $p$ and $q$ scales as $\sqrt{\vert p-q\vert}$, as expected from the central limit theorem. 

The stochastic process $\ve[R](t)$ is Gaussian, its dynamics can therefore be characterized by the evolution of its average and variance. Let us consider initial conditions where the polymer is at equilibrium, with the supplementary condition that $\ve[R]=\ve[R]_0$. Then, the average of $\ve[R](t)$ at a later time is given by $\langle\ve[R](t)\rangle=\phi(t)\ve[R]_0$: the function $\phi(t)$ describes how the average of $\ve[R](t)$ relaxes to its equilibrium position. This function is such that $\phi(0)=1$, and $\phi(t)$ vanishes at large times. The function $\phi$ is given by the following expression, whose derivation is reminded in Appendix \ref{AppendixFunctionPhiAndPsi}:
\begin{align}
	\phi(t)=\sum_{i=2}^N \frac{b_i^2 e^{-\lambda_i t}}{\lambda_i L^2}\label{DefinitionPhi}.
\end{align}
We also define a function $\psi(t)$, that  represents the variance of each coordinate of $\ve[R](t)$, given that initially the value of $\ve[R]$ is $\ve[R]_0$ and that the polymer is at equilibrium:
\begin{align}
	\psi(t)= L^2[1-\phi(t)^2]\label{DefinitionPsi}.
\end{align}
From the expressions (\ref{DefinitionPhi}),(\ref{DefinitionPsi}), we deduce the behavior of the mean square displacement function $\psi(t)$ at different time scales when $N\gg1$:
\begin{align}
\psi(t)\simeq 
\begin{cases}
4  t  & (t\ll 1 \ ; \ \Delta R \ll 1) \\
\kappa \sqrt{t} & (1\ll t \ll N^2 \ ; \ 1\ll \Delta R\ll \sqrt{N}) \\
L^2 & (t \gg N^2 \ ; \ \Delta R\simeq \sqrt{N})
\end{cases} \label{ComportementPsi}.
\end{align}
In Eq. (\ref{ComportementPsi}), the behaviors are distinguished for the different time scales $t$. The limiting behaviors can also be discussed for the different length scales $\Delta R\sim\psi^{1/2}$ as well. We remind that the length $l_0$ and the relaxation time $\tau_0$ have been chosen as units of length and time. 
According to (\ref{ComportementPsi}), the motion is diffusive at short times, where the reactive monomers behave as if they were disconnected from the rest of the chain. At intermediate time scales, $\psi\sim \kappa\ t^{1/2}$, the motion is subdiffusive and results from the interactions with all the monomers of the chain. The coefficient $\kappa$ does not depend on $N$, but depends on the position of the reactive monomer in the chain ($\kappa=8/\sqrt{\pi}$ if the reactive monomers are at the chain ends, whereas $\kappa=4/\sqrt{\pi}$ for interior reactive monomers, see Appendix \ref{AppendixFunctionPhiAndPsi}). At long times, $\psi$ is constant, and the process $\ve[R]$ reaches the stationary state. 

From the behavior (\ref{ComportementPsi}) of the mean square displacement $\psi$ at different time scales, we can derive a simple scaling law for the reaction time by using the systematic procedure introduced in a recent paper dealing with intermolecular reactions\cite{guerin2012b}. 
Note that alternative qualitative reasonings can be found in other references\cite{Toan2008,Chen2005}.
The fact that $\psi\rightarrow L^2$ for large times indicates that the parameter $L^3\simeq N^{3/2}$ plays the role of an effective  confining volume. Let us assume that the size of the reactive region is small compared to the length of a single bond ($a\ll 1$), and that initially the polymer is at equilibrium. 
The diffusion controlled reaction occurs in two sub-steps that involve the properties of the stochastic process $\ve[R](t)$ at two different length scales. 
The first step of the reaction consists in reaching for the first time a distance of order $1$ between the reactants. According to (\ref{ComportementPsi}), at these large length scales, $\ve[R]$ is subdiffusive:  $\psi\sim\langle \ve[R](t)^2\rangle\sim t^{1/2}=t^{2/d_w}$, defining a walk dimension $d_w=4$. 
A Markovian walker that has a walk dimension of $d_w$ and starts in a random position in a confining volume of radius $L$ would reach a punctual target in a time\cite{Condamin2007,Tejedor2009,benAvraham2000} $T\simeq L^{d_w}$. Assuming that this scaling argument holds for our non-Markovian problem, we deduce that 
the average time needed to complete the first step of the reaction scales as $T\simeq L^{d_w}\sim N^2$. 

Once the length $l_0=1$ has been reached for the first time, there is a second step in the reaction, that consists in reaching the sphere of radius $a$. From (\ref{ComportementPsi}), we observe that, at small length scales, $\ve[R](t)$ behaves as a diffusive process with diffusion coefficient of oder $1$. Assuming that the reaction time is the same than for a diffusive Markovian walker\cite{Benichou2008,Singer2006a} in a confining volume $V=L^3$ with an initial position that is far from the reactive site, we get the estimate for the time of this reaction substep: $T\simeq L^3/a$. Adding the two times corresponding to the two substeps, we obtain the total reaction time:
\begin{align}
	T \simeq \  \frac{N^{3/2}}{a} + N^2 \label{ScalingRelationQualitative}.
\end{align}
From (\ref{ScalingRelationQualitative}), it is clear that the dominant term for the reaction time is $N^2$ when $N$ is large and comes from the subdiffusive reaction substep. However, the first term of (\ref{ScalingRelationQualitative}) becomes important when the size $a$ of the reactive region is smaller than $N^{-1/2}$.  

The presence of the regimes appearing in Eq. (\ref{ScalingRelationQualitative}) for the reaction time is already known \cite{Toan2008,Chen2005,DOI1975,Pastor1996}. 
The earliest discussion was done by Doi\cite{DOI1975}, who noted that the scaling $T\sim N^2$, that is predicted by the Wilemski-Fixman theory \cite{WILEMSKI1974a,WILEMSKI1974b} for large $N$, appears only when the interactions between the monomers are considered. Indeed, the fact of replacing the whole polymer by a single spring leads to the different law $T\simeq N^{3/2}a^{-1}$ for the reaction time\cite{SUNAGAWA1975}, a result which was recovered with another approach of the harmonic spring model in the SSS theory of Szabo, Schulten and Schulten\cite{Szabo1980}.  
Importantly, as noted in a previous work\cite{Pastor1996} and reminded below, the Wilemski-Fixman treatment of the full problem predicts both behaviors $N^2$
 and $N^{3/2}a^{-1}$ that appear in the large $N$ and small $a$ limits, respectively. The behavior $N^{3/2}a^{-1}$ has also been recently derived from first principles\cite{Amitai2012}, while the scaling relation $T\sim N^2$  appears in the treatment of the problem by the renormalization group theory\cite{Friedman1988}. The presence of the two regimes of 
(\ref{ScalingRelationQualitative}) has  been checked with numerical simulations only recently\cite{Chen2005}. 
At this stage, we have used a simple scaling argument to derive the scaling relation  (\ref{ScalingRelationQualitative}), where the apparition of two regimes is linked to the presence of two substeps where the motion of the reactant is qualitatively different, corresponding to the different regimes appearing in Eq. (\ref{ComportementPsi}). Note that the scaling relation (\ref{ScalingRelationQualitative}) is not sufficient to characterize the  reaction time, as it does not describe its behavior for finite values of $N$ and $a$ and it does not permit the identification of the numerical coefficients. In the next sections, we describe a non-Markovian theory that enables the precise derivation of the reaction time. We also discuss the validity of the hypotheses made in the existing theories in the context of our more general theory.  

\section{Non-Markovian theory for the kinetics of intramolecular reactions}
\label{SectionNonMarkovianTheory}
We now derive the equations of the non-Markovian theory of intramolecular reaction kinetics that was briefly introduced in a previous work\cite{Guerin2012}. As these equations share a lot of similarities with the equations for intramolecular reactions that are described in details in Ref. \cite{guerin2012b}, we refer to this reference and to Appendix \ref{AppendixDerivationEquationMPi} for  calculation details. 
The starting point of our analysis is to consider the stochastic process formed by the positions of all the monomers  $\ve[r]_i(t)$ (or, equivalently, of all the modes $\ve[a]_i$). This process, to the difference of 
the single process $\ve[R](t)$, is Markovian, which enables us to use the renewal theory\cite{VanKampen1992}.  
Let us temporarily consider the case where the polymer does not react when it reaches the reactive zone. We define an arbitrary position $\ve[R]_f$ that is inside the reactive zone (\textit{i.e.} with $\|\ve[R]_f\|\le a$), and we consider a configuration $\vert \ve[a]\rangle$ such that $\ve[R]=\ve[R]_f$. If this configuration is observed at $t$, it means that the polymer must have crossed the reactive boundary at some earlier time $t'$, with a configuration $\vert \ve[a]'\rangle$. Therefore, defining $f(\vert\ve[a]'\rangle,t')$ as the probability density of reacting for the first time at $t'$ with a configuration $\vert\ve[a]'\rangle$, we can write the following renewal equation\cite{VanKampen1992}:
\begin{equation}
P(\vert \ve[a]\rangle,t\vert \{\text{ini}\})=\int_0^t dt' \int d\vert \ve[a]'\rangle f(\vert \ve[a]'\rangle,t')P(\vert \ve[a]\rangle,t \vert \ \vert \ve[a]'\rangle,t') \label{RenewalEquation}.
\end{equation}
In this equation, $d\vert\ve[a]\rangle=d\ve[a]_1...d\ve[a]_N$, and $P(\vert \ve[a]\rangle,t \vert \ \vert \ve[a]'\rangle,t')$ is the probability of observing the configuration $\vert\ve[a]\rangle$ at $t$ given that the configuration $\vert\ve[a]'\rangle$ was observed at time $t'$. Similarly, $P(\vert \ve[a]\rangle,t\vert \{\text{ini}\})$ is the probability of observing the configuration $\vert\ve[a]\rangle$ at $t$ given an initial distribution of conformations at $t=0$. The initial distribution of modes is assumed to be an equilibrium distribution, given that the initial distance between the reactants is larger than the reactive radius ($R_0>a$). Let us introduce $\Omega$, which is  a shortcut for  both the polar angle $\theta$ and the azimuthal angle $\varphi$, with $d\Omega=\sin\theta d\theta d\varphi/4\pi$, and $\ve[u]_r(\Omega)$ the unit radial vector pointing outwards the unit sphere in the direction defined by $(\theta,\varphi)$. Then, the initial distribution of conformations can be written as: 
\begin{align}
&P_{\text{ini}}(\vert \ve[a]\rangle)=\nonumber\\
&\int_a^{\infty} \frac{dR_0 R_0^2 e^{-R_0^2/(2L^2)}}{Z(a,L^2)}\int d\Omega P_{\text{stat}}(\vert \ve[a] \rangle\vert R_0 \ve[u]_r (\Omega) ), \label{DefinitionAverageOverInitialCondition}
\end{align}
where $P_{\text{stat}}(\vert \ve[a] \rangle \vert \ve[R]_1)$ represents the equilibrium probability density of conformations given that $\ve[R]=\ve[R]_1$, and  where the normalization factor $Z$ is defined by:
\begin{align}
	Z(a,h)=\int_a^{\infty}dR_0 \ R_0^2 \  e^{-R_0^2/(2h)} \label{DefinitionZ}.
\end{align}
Taking the Laplace transform of Eq. (\ref{RenewalEquation}), and developing for small values of the Laplace variable leads to an equation that involves the mean first passage time:
\begin{align}
	&T P_{\text{stat}}(\vert \ve[a]\rangle)=\nonumber\\
	&\int_0^{\infty}dt \int d\Omega \Big[P(\vert \ve[a]\rangle,t\vert \pi_{\Omega},0)-P(\vert \ve[a]\rangle,t\vert P_{\text{ini},\Omega},0)\Big].\label{EquationIntegraleSplitting}
\end{align}
In Eq. (\ref{EquationIntegraleSplitting}), we have introduced the splitting probability $\pi_{\Omega}(\vert \ve[a]\rangle)$, which represents the distribution of configurations $\vert \ve[a]\rangle$ at the very instant of reaction, given that $\ve[R]=a\ \ve[u]_r(\Omega)$ at the instant of reaction. Similarly, $P_{\text{ini},\Omega}(\vert\ve[a]\rangle)$ represents the initial distribution of modes given that $\ve[R]$ is in the direction $\ve[u]_r(\Omega)$. The quantity $P(\vert \ve[a]\rangle,t\vert \pi_{\Omega},0)$ represents the probability density of observing $\vert \ve[a]\rangle$ at $t$ given that the initial distribution of the modes was $\pi_{\Omega}$. Equation (\ref{EquationIntegraleSplitting}) is an exact integral equation that defines both the reaction time $T$ and the splitting probability distribution $\pi_{\Omega}$. Integrating it over the configurations that are such that $\ve[R]=\ve[R]_f$ leads to the following exact expression of the reaction time:
\begin{align}
	&T P_{\text{stat}}(\ve[R]_f)=\nonumber\\
	&\int_0^{\infty}dt \int d\Omega \Big[P(\ve[R]_f,t\vert \pi_{\Omega},0)\nonumber\\
	&-\frac{1}{Z(a,L^2)}\int_a^{\infty}dR_0 R_0^2 e^{-\frac{R_0^2}{2L^2}} P(\ve[R]_f,t\vert \{R_0\ve[u]_r,\text{stat}\},0)\Big],	\label{EstimationMFPT_splitting_dim3_ExpliciteAverage}
\end{align}
in which $P(\ve[R]_f,t\vert \pi_{\Omega},0)$ represents the probability density of observing $\ve[R]$ in position $\ve[R]_f$, given the initial distribution $\pi_{\Omega}$ for the conformations  at time $t=0$. 
Equation (\ref{EstimationMFPT_splitting_dim3_ExpliciteAverage}) shows that the reaction time is inversely proportional to the factor $P_{\text{stat}}(\ve[R]_f)$, which is the  equilibrium probability density to find the two reactive monomers in contact, and is sometimes referred to as the $j$-factor\cite{Allemand2006,Rippe2001}. It plays the same role as the inverse of a confinement volume in the case of intermolecular reactions\cite{guerin2012b}, and it can be written explicitly:
\begin{align}
	P_{\text{stat}}(\ve[R]_f)=\frac{1}{(2\pi L^2)^{3/2}} \text{exp}\left(-\frac{\ve[R]_f^2}{2L^2}\right).
\end{align}
Equation (\ref{EstimationMFPT_splitting_dim3_ExpliciteAverage}) is an exact expression of the reaction time. However, it cannot be used without knowing the splitting probability distribution $\pi_{\Omega}$.  

As in the case of intermolecular reactions\cite{guerin2012b}, the key hypothesis of the non-Markovian theory is to assume that the splitting distribution $\pi_{\Omega}$ is a multivariate Gaussian distribution: only the values of the average and the covariance of the modes $\ve[a]_i$ over $\pi_{\Omega}$ have to be determined. This Gaussian approximation of $\pi_{\Omega}$ considerably simplifies the problem. Instead of having to find a function $\pi_{\Omega}(\ve[a]_1,...,\ve[a]_N)$ of $3N$ variables that is solution of the integral equation (\ref{EquationIntegraleSplitting}), one is left to finding a finite set of unknown quantities, that are the first and second moments of $\pi_{\Omega}$. 

The first moments of $\pi_{\Omega}$ are the average value of the modes $\ve[a]_i$ at the instant of the reaction, given that the reaction takes place in the direction $\Omega$ and are noted $\ve[m]_i^{\pi}$. For symmetry reasons, $\ve[m]_i^{\pi}$ is oriented in the radial direction defined by the unit vector $\ve[u]_r(\Omega)$, and we call $m_i^{\pi}$ its component in this direction: $\ve[m]_i^{\pi}=m_i^{\pi}\ve[u]_r$. 
At a time $t$ after the reaction, the average value of $\ve[a]_i$ in this direction is simply $m_i^{\pi}e^{-\lambda_i t}$. Summing over the coefficients $b_i$, we deduce the value of $R_{\pi}(t)$, defined as the average value of $\ve[R]$ in the radial direction $\ve[u]_r(\Omega)$ at a time $t$ after the reaction, which reads:
\begin{align}
	R_{\pi}(t)=\sum_{i=2}^{N}b_i m_i^{\pi} e^{-\lambda_i t}.\label{DefinitionRpi}
\end{align}

We will derive below a set of equations that define the moments $m_i^{\pi}$ in a self-consistent way. In principle, in the non-Markovian theory, one should also determine the covariance matrix of $\pi_{\Omega}$. 
Here, for simplicity, we assume that the covariance matrix of $\pi_{\Omega}$ is well approximated by the covariance matrix that characterizes the equilibrium conformations that make a loop. We call this approximation the ``stationary covariance approximation'', which was found to be accurate in the case of intermolecular reactions\cite{Guerin2012,guerin2012b}. We deduce the value of the equilibrium covariance matrix from the formulas on conditional Gaussian probability distributions that are mentioned in  Appendix \ref{AppendixProjectionPropagationFormulas} and Appendix
 \ref{AppendixFunctionPhiAndPsi}: 
\begin{align}
\text{cov}(a_{i,\alpha}\ , \ a_{j,\beta})=\left(\frac{\delta_{ij}}{\lambda_i}-\frac{b_i b_j }{\lambda_i \lambda_j L^2}\right)\delta_{\alpha\beta}\label{StationaryCovarianceApprox},
\end{align}
where $\alpha$ and $\beta$ represent the spatial coordinates $x,y,z$. Note that, although the modes $\ve[a]_i$ are independent at equilibrium, the fact of conditioning them to a particular value of  the vector $\ve[R]=\sum b_i \ve[a]_i$ introduces correlations between them. In the stationary covariance approximation, the moments $m_i^{\pi}$, together with the reaction time $T$, are the only unknown variables of the theory.

Under the stationary covariance approximation, we can  write an explicit form of the reaction time (\ref{EstimationMFPT_splitting_dim3_ExpliciteAverage}): 
\begin{align}
	T P_{\text{stat}}&(\ve[R]_f)=	\int_0^{\infty}dt \int \frac{d\Omega}{(2\pi\psi)^{3/2}}\Bigg[ e^{-\frac{[\ve[R]_f-R_{\pi}\ve[u]_r(\Omega)]^2}{2\psi}}\nonumber\\
	&-\frac{1}{Z(a,L^2)} \int_a^{\infty} dR_0 e^{-\frac{R_0^2}{2L^2}} e^{-\frac{[\ve[R]_f-R_{0}\phi\ve[u]_r(\Omega)]^2}{2\psi}}\Bigg]\label{74830},
\end{align}
where the functions $\phi$ and $\psi$ describe the relaxation dynamics towards equilibrium, and have been introduced previously in Eqs. (\ref{DefinitionPhi}),(\ref{DefinitionPsi}). The integral over the angles $\Omega=\theta,\phi$ in (\ref{74830}) can be performed by noting that $(\ve[R]_f-R_{\pi}\ve[u]_r)^2=R_f^2+R_{\pi}^2-2R_f R_{\pi}\cos\theta$, but the resulting expression is however not simple. At this stage, we note that, by construction, the theory should predict the same value of $T$ whatever the choice of the final position $\ve[R]_f$. In particular, with $\ve[R]_f=\ve[0]$, we obtain a simpler expression for the reaction time, which writes:
\begin{align}
	&\frac{T}{ L^3}=	&\int_0^{\infty} \frac{dt}{\psi^{3/2} } \left[\exp \left(-\frac{R_{\pi}^2}{2\psi} \right)-\frac{Z(a,\psi)}{Z(a,L^2)}\right].	\label{EstimationMFPT_splitting_dim3_Explicite_Centered}
\end{align}
where we remind that the function $Z$ is defined in Eq. (\ref{DefinitionZ}). This important expression is the simplest expression of the reaction time in the non-Markovian theory.

At this stage, we need to write a set of equations that enable the calculation of the moments $m_i^{\pi}$. This set of equations is found by multiplying the general integral equation (\ref{EquationIntegraleSplitting}) by the factor $\ve[a]_i \delta(\ve[R]_f-\sum_j b_j \ve[a]_j)$, and integrating over all the modes (see Appendix \ref{AppendixDerivationEquationMPi}). Because of the presence of the $\delta$ function, the resulting terms involve the conditional average $\mu_i^{\pi,0}$, which represents the average value of the $i^{\text{th}}$  mode at a time $t$ after the reaction, given that the vector $\ve[R]$ has a value $\ve[R]=\ve[0]$ at the same time $t$. The expression of $\mu_i^{\pi,0}$ can be deduced from the formulas on conditional Gaussian distributions that are given in Appendix \ref{AppendixProjectionPropagationFormulas}:
\begin{align}
	\mu_i^{\pi,0}=m_i^{\pi}e^{-\lambda_i t} - \frac{R_{\pi}(t) b_i[1-\phi(t)e^{-\lambda_i t}]}{\lambda_i\psi(t)}.
\end{align}
After the multiplication of the integral equation (\ref{EquationIntegraleSplitting}) by the factor $\ve[a]_i \delta(\ve[R]_f-\sum_j b_j \ve[a]_j)$, and integrating over all the modes leads to to the following set of self-consistent equations, that are derived in detail in Appendix \ref{AppendixDerivationEquationMPi}:
\begin{align}
&\int_0^{\infty}\frac{dt}{\psi^{5/2} } \Bigg\{ \left[\frac{\mu_{i}^{\pi,0}R_{\pi}}{3}+ 
\frac{b_i\phi (\phi- e^{-\lambda_i t})}{\lambda_i } \right]e^{-\frac{R_{\pi}^2}{2\psi}} \nonumber\\
&- \frac{b_i \phi (\phi-e^{-\lambda_i t}) }{Z(a,L^2) \lambda_i}  \left[Z(a,\psi)-\frac{G(a,\psi)}{3\psi}
\right] \Bigg\} =0,\label{EquationFirstMomentDim3AveragedSimplified}
\end{align}
where we have defined the function $G$  by the following relation:  
\begin{align}
G(a,h)=\int_a^{\infty}dR_0 R_0^4 \exp\left(-\frac{R_0^2}{2h}\right)\label{DefinitionFunctionG}.
\end{align}
As already mentioned, the conditional average $\mu_i^{\pi,0}(t)$ plays a key role in the self-consistent equations that define $m_i^{\pi}$. Other terms come from the results of the average over angles $\Omega$ and initial distance $R_0$. The relation (\ref{EquationFirstMomentDim3AveragedSimplified}) is valid for $2\le i\le N$ and actually forms a set of $N-1$ equations that entirely define the moments $m_i^{\pi}$ ; it is the key equation of the non-Markovian theory as it enables to define in a self-consistent way the average value of the polymer conformation at the instant of the reaction. In fact, there are only $N-2$ equations, because $\ve[R]$ is known to be equal to $a\ \ve[u]_r$ with probability one over the distribution $\pi_{\Omega}$, which implies the relation $\sum_j b_jm_j^{\pi}=a$ that is compatible with (\ref{EquationFirstMomentDim3AveragedSimplified}). 
Solving the non-Markovian theory consists in solving the set of equations (\ref{EquationFirstMomentDim3AveragedSimplified}) to obtain the average value of the modes $a_i$  at the reaction, which can be done numerically or analytically in some limiting cases (see below). Then, the result is inserted into the expression of the reaction time (\ref{EstimationMFPT_splitting_dim3_Explicite_Centered}). We stress that, until now, we did not make any hypothesis on the actual location $p$ and $q$ of the reactive monomers, which enter only in the definition (\ref{ExpressionCoefficient}) of the coefficients $b_i$ and therefore in the functions $\phi$ and $\psi$. 

\section{Relation of the non-Markovian theory to existing Markovian theories of cyclization}
\label{SectionComparisonTheories}

Now, we show how introducing supplementary approximations in the non-Markovian theory enables to recover some existing theories. All the theories described in this section will be confronted to simulations in section \ref{SectionComparisonTheoriesWithSimus}. For simplicity, we consider the case where the reactive monomers are at the polymer extremities ($p=1$ and $q=N$). 

The first theory that we consider is the Wilemski-Fixman theory\cite{WILEMSKI1974b,WILEMSKI1974a}, which turns out to appear naturally in our equations by making a local equilibrium assumption. 
A Markovian approximation consists in assuming that the distribution of conformations at the instant of reaction is an equilibrium distribution, conditional to the restriction that $\ve[R]=a \ \ve[u]_r(\Omega)$, where $\ve[u]_r(\Omega)$ is the unit vector defining the direction in which the reaction takes place. Formally, this approximation can be written as:
\begin{align}
	\pi_{\Omega}(\vert \ve[a]\rangle)\simeq P_{\text{stat}}(\vert \ve[a]\rangle \vert a\ \ve[u]_r(\Omega) )\label{MarkovianApproximationWriting}.
\end{align}
In our theory, we already assumed that the covariance matrix of $\pi_{\Omega}$ is given by the stationary covariance matrix, so that the approximation (\ref{MarkovianApproximationWriting}) consists in assuming that the moments $m_i^{\pi}$, instead of satisfying the self-consistent equations (\ref{EquationFirstMomentDim3AveragedSimplified}), are equal to the average value of the modes at equilibrium, given the condition $\ve[R]=a\ve[u]_r$. Applying a formula on conditional Gaussian distributions that is reminded in Appendix \ref{AppendixProjectionPropagationFormulas}, we find that this approximation leads to the following expression for the the average value of the modes $m_i^{\pi}$:
\begin{align}
	m_i^{\pi}\simeq\frac{a \ b_i}{L^2\lambda_i} \label{Value_of_mq_Markovian}.
\end{align}
We can gain insight in the meaning of this formula by considering $\langle z_i\rangle_{\pi}=\sum_j Q_{ij}m_j^{\pi}$, which is the average value of the position of the $i^{\text{th}}$ monomer in the radial direction at the instant of reaction.  Approximation (\ref{Value_of_mq_Markovian}) implies that $\langle z_i\rangle_{\pi}$ is linear in the part of the polymer between the reactive groups ($p\le i\le q$), whereas it is constant at the exterior. 

Inserting (\ref{Value_of_mq_Markovian}) into (\ref{DefinitionRpi}) and comparing with  (\ref{DefinitionPhi}) implies that the reactive trajectory in the Markovian approximation is simply $R_{\pi}(t)\simeq \ a \ \phi(t)$. Then, evaluating the general formula (\ref{EstimationMFPT_splitting_dim3_ExpliciteAverage}) with these approximations and taking the value $R_f=a$, we obtain the following expression for the reaction time in the Markovian approximation: 
\begin{align}
	T\simeq	
\int_0^{\infty} dt \left\{\frac{L^2 e^{-\frac{a^2\phi^2}{L^2(1-\phi^2)}}}{ a^2 \phi(1-\phi^2)^{1/2} }\text{sinh}\left[\frac{a^2\phi}{L^2(1-\phi^2)}\right]
-1\right\}\label{ExpressionTMarkovianExt}.
\end{align}
In fact, the expression (\ref{ExpressionTMarkovianExt}) is valid only when $a\ll L$, otherwise the term ``$-1$'' in the second part of the integrand would be replaced by a more complicated term (strictly speaking, (\ref{ExpressionTMarkovianExt}) gives the mean first passage time to the reactive sphere, but with an initial distribution that is an equilibrium distribution which can be inside or outside the reactive region).
The expression (\ref{ExpressionTMarkovianExt}) is the reaction time obtained by using a Wilemski-Fixman approach and choosing a delta sink function\cite{Pastor1996}, which is the most accurate choice of sink function to describe reactions in the diffusion controlled regime\cite{Pastor1996}. This result clearly shows that the Wilemski-Fixman approach is equivalent to the assumption that the reactive conformations of the polymer can be replaced by equilibrium looping conformations. 
Our theory also provides an alternative formula for the reaction time, that is simpler than  (\ref{ExpressionTMarkovianExt}), and that is obtained by calculating (\ref{EstimationMFPT_splitting_dim3_ExpliciteAverage}) with the value $R_f=0$:
\begin{align}
	\frac{T_{\text{Mark.}}}{L^3}=\int_0^{\infty} \frac{dt}{\psi(t)^{3/2}} \left[e^{-a^2\phi^2/(2\psi)}-\frac{Z(a,\psi)}{Z(a,L^2)}\right]\label{ExpressionTMarkovianCentre}.
\end{align}
The two formulas (\ref{ExpressionTMarkovianExt}),(\ref{ExpressionTMarkovianCentre}) are two legitimate expressions of the reaction time in the Markovian approximation. 

Let us briefly derive the scaling relations predicted by the Markovian theory from Eq. (\ref{ExpressionTMarkovianCentre}) in the case of the cyclization  reaction. First, let us consider the limit of a large number of monomers, at a fixed value of $a$. In this limit, the function $\phi$ tends to a limiting function $\Phi(\tau)$ that depends on the rescaled time $\tau=t/N^2$, and that is deduced from (\ref{DefinitionMatrixQ}),(\ref{ExpressionCoefficient}),(\ref{DefinitionPhi}):
\begin{align}
\Phi(\tau)=\lim_{N\rightarrow\infty} \phi(\tau N^2)=\sum_{q  \ \text{odd}} \frac{8}{q^2\pi^2}e^{-\pi^2 q^2\tau}\label{DefinitionRescaledPhi}.
\end{align}
Due to the infinite number of terms in (\ref{DefinitionRescaledPhi}), the function $\Phi$ has an anomalous behavior for small values of the rescaled time: $\Phi(\tau)\simeq 1-4(\tau/\pi)^{1/2}$, which reflects the subdiffusive behavior of the end-to-end vector. Due to this anomalous behavior, we can rescale (\ref{ExpressionTMarkovianCentre}) and express the reaction time as a form of a convergent integral:
\begin{align}
T&\simeq N^2 \int_0^{\infty} d\tau \left\{\frac{1}{[1-\Phi(\tau)^2\ ]^{3/2}}-1\right\} \ (N\rightarrow\infty)  \label{MarkovianScalingLargeN}.
\end{align}
Evaluating numerically this integral leads to the formula $T\simeq 0.201 \ N^2$: we recover the fact that, in the Wilemski-Fixman theory, the mean cyclization time is proportional to $N^2$ for large $N$ at fixed $a$. It is however interesting to note that the limit of small reactive region leads to  another scaling relation. In the limit $a\rightarrow0$ at fixed $N$, the expression (\ref{ExpressionTMarkovianCentre}) can be evaluated by replacing the integrand by its short time limit (in which $\phi\simeq1$ and $\psi\simeq 4 t$):
\begin{align}
T\simeq L^3\int_0^{\infty} dt\frac{e^{-a^2/(8t)}}{(4\ t)^{3/2}} = \frac{ \sqrt{\pi} \ (N-1)^{3/2}}{\sqrt{8}\ a} \ (a\rightarrow0)\label{MarkovianScalingSmall_a}.
\end{align}
Therefore, the Markovian theory contains the two scalings $T\sim N^{3/2}/a$ and $T\sim N^2$, a fact that had already been noted by Pastor \textit{et al.}\cite{Pastor1996}. The two scalings (\ref{MarkovianScalingLargeN}),(\ref{MarkovianScalingSmall_a}) correspond exactly to the general scaling relation (\ref{ScalingRelationQualitative}), that was derived by assuming that the reaction occurs in two substeps, that involve the subdiffusive and diffusive behavior of the end-to-end vector at large and intermediate time scales, respectively.

It turns out that the formula (\ref{MarkovianScalingLargeN}) can be linked with the results of the renormalization group theory of cyclization, that are valid when the value of the spatial dimension $d$ is close to $d=4$. A simple generalization of (\ref{MarkovianScalingLargeN}) to the case of a $d$-dimensional space leads to:
\begin{align}
	T = N^2 \int_0^{\infty} d\tau \left\{\frac{1}{[1-\Phi(\tau)^2]^{d/2}}-1\right\}\label{GeneralizedDimensionExpressionMFPT}.
\end{align}
This integral converges only for $d<4$. If we introduce a fixed (small) time $t_1$, we see that in the  limit $4-d\rightarrow0$, (\ref{GeneralizedDimensionExpressionMFPT}) is approximately equal to:
\begin{align}
	T \simeq N^2 \int_0^{t_1} d\tau \frac{1}{[8/\sqrt{\pi} \tau^{1/2} ]^{[2-(4-d)/2]}},  \label{Integral} 
\end{align}
where we neglect supplementary terms that do not diverge as $d\rightarrow4$. By performing the integral (\ref{Integral}), we find that the result at smallest order in the parameter $4-d$ is: 
\begin{align}
	T\simeq \frac{ \pi \ N^2}{16 (4-d)}=T_{\text{RG}} \hspace{1cm} (d\rightarrow4,N\rightarrow\infty). \label{ResultRenormalizationGroup} 
\end{align}
The expression (\ref{ResultRenormalizationGroup}) is exactly the scaling relation obtained in the renormalization group theory\cite{Friedman1988}. Interestingly, setting $d=3$, we find that the 
 the numerical coefficient of the scaling relation (\ref{ResultRenormalizationGroup}) is $\pi/16= 0.196$. This value differs by only $2.4\%$ from the value of $0.201$ that was estimated by the Markovian approximation. 
The fact that this result of the renormalization group can be derived by developing the reaction time obtained from the Markovian theory suggests that there is a local equilibrium assumption that is made in the renormalization group theory.

Finally, we discuss the evaluation of the mean first passage time in the simplest theory of cyclization: the harmonic spring approximation, which consists in replacing the whole polymer chain by a single spring that has an effective stiffness $k_{\text{eff}}$. Due to the fact that there are $N-1$ bonds in series, this effective stiffness is $k_{\text{eff}}=k/(N-1)$. Then, one assigns a single effective drag coefficient $\zeta_{\text{eff}}$ (which is to be determined later) to the end-to-end vector $\ve[R]$. With these assumptions, the process $\ve[R]$ is now characterized by a single time scale $\tau_{\text{eff}}=\zeta_{\text{eff}}/k_{\text{eff}}$. It is therefore Markovian and its first passage properties can therefore be computed analytically, as was done by Szabo, Schulten and Schulten\cite{Szabo1980} who determined the reaction time by directly solving the adjoint equation and who obtained:
\begin{align}
	T_{\text{SSS}}= \frac{\tau_{\text{eff}} }{L^2 Z(a,L^2)}\int_a^{\infty} \frac{dx
}{x^2 e^{-x^2/(2L^2)}} \left[\int_x^{\infty}dy\  y^2 e^{-\frac{y^2}{2L^2}}\right]^2 \label{TempsReactionSSSGlobal}.
\end{align}
An alternative method to solve the harmonic spring problem is to use the Renewal theory, whose result is directly given by equation (\ref{EstimationMFPT_splitting_dim3_Explicite_Centered}), in which $\phi$ must be replaced by its value for an Ornstein-Uhlenbeck process ($\phi=\exp(-t/\tau_{\text{eff}})$). 
Of course, these two methods lead to the same results, as they both are an exact treatment of the harmonic spring model. If we take the limit $a\rightarrow0$ or $N\rightarrow\infty$ in the SSS expression (\ref{TempsReactionSSSGlobal}), we obtain at leading order: 
\begin{align}
	T_{\text{SSS}} \simeq \frac{ \sqrt{\pi} \ \tau_{\text{eff}}\ \sqrt{N-1} }{\sqrt{2} \ a} \hspace{0.5cm} (a\ll \sqrt{N-1}) \label{TempsReactionSSS_Small_a}.
\end{align}
In the harmonic spring approximation, the choice of the effective drag coefficient $\zeta_{\text{eff}}$ is quite arbitrary. A natural choice is to set $\zeta_{\text{eff}}=\zeta/2$, which corresponds to the approximation that the two end-monomers behave as if they were disconnected from the rest of the chain. With this choice, the scaling relation  (\ref{TempsReactionSSS_Small_a}) coincides with the Markovian scaling law (\ref{MarkovianScalingSmall_a}) obtained for small $a$. 
However in this case, the effective relaxation time $\tau_{\text{eff}}=\zeta (N-1)/(2k)$ is very different from the largest relaxation  time of the polymer $\tau_R=N^2\zeta/(\pi^2 k)$, which is one of the reasons why the SSS theory is not expected to give an accurate estimation of the reaction time. 

In the next section, we compare all these different theories with the results of numerical stochastic simulations.

\section{Comparison of the different theories of cyclization with numerical simulations}
\label{SectionComparisonTheoriesWithSimus}

\begin{figure}[h!]
\includegraphics[width=7.5cm,clip]{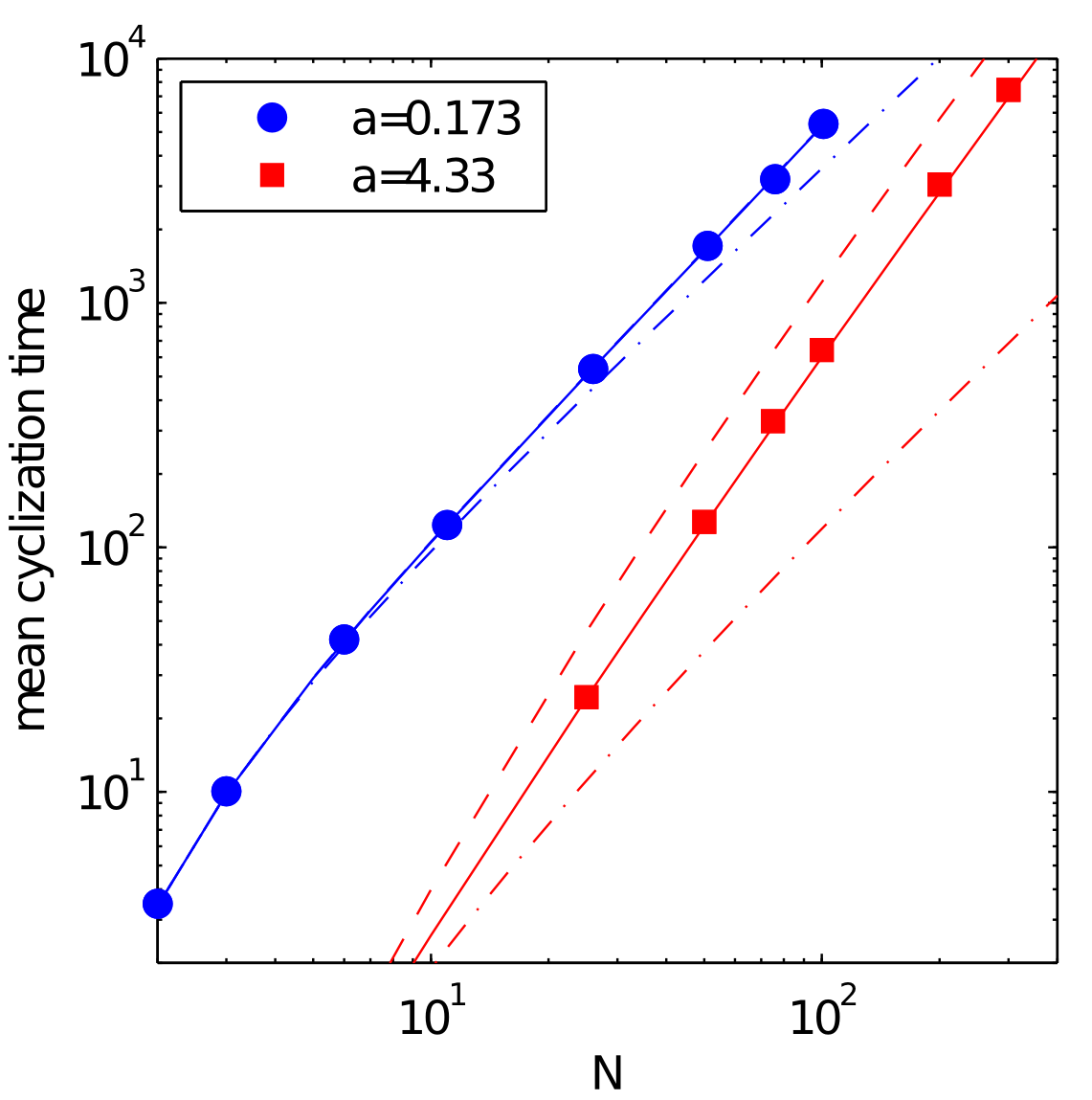}
\caption{Average end-to-end cyclization time in simulations (symbols) and different theories (lines). There are 2 sets of curves, corresponding to 2 values of $a$ (upper blue curves and circles: $a=0.1\sqrt{3}$, lower red curves and squares: $a=2.5\sqrt{3}$). In each set of curves, the continuous lines are not a fit but represent the results of the non-Markovian theory. The upper dashed lines represent the results of the Wilemski-Fixman Markovian theory evaluated with equation (\ref{ExpressionTMarkovianCentre}), while the lower dash-dot lines represent the results of the SSS theory evaluated with equation (\ref{TempsReactionSSSGlobal}). For the smallest value of $a$, the non-Markovian and the Markovian results are undistinguishable on this figure.
\label{FigureComparaisonSimuAvecSSS}}
\end{figure}  

\begin{table*} 
\caption{\label{TableauResultatsSimulationsTheories} Values of the end-to-end cyclization time determined by stochastic simulations and different theories. $T_{\text{simulation}}$ is the average first cyclization time in stochastic simulations, with $\delta T$ giving the $95\%$ confidence interval resulting from the error due to the finite number of simulation runs. 
$T_{\text{Non-Markovian}}$ is the result of the non-Markovian theory presented in this paper, $T_{\text{Markovian}}$ is the result of the Markovian (Wilemski-Fixman\cite{WILEMSKI1974b,WILEMSKI1974a}) approximation and is evaluated with (\ref{ExpressionTMarkovianCentre}).  $T_{\text{SSS}}$ is the result of the SSS harmonic spring approximation\cite{Szabo1980} and is given by the formula (\ref{TempsReactionSSSGlobal}). $T_{\text{RG}}$ is the result of the renormalization group approach and is given by formula (\ref{ResultRenormalizationGroup}). Last,  $T_{\text{AKH}}$ is the result of Amitai, Kupka and Holcman\cite{Amitai2012} and is given by $T_{\text{AKH}}=\sqrt{\pi/8}N^{3/2}/a+3A_3N^2$, with $A_3\simeq0.05$. $T_{\text{RG}}$ and $T_{\text{AKH}}$ are only asymptotically valid for small $N$ and $a$, respectively, and are shown here only for the sake of completion. The simulation results for $a=0.87$ are taken from Pastor \textit{et al.}\cite{Pastor1996}. 
Note that $a$ is in units of $l_0$ and that all the times are expressed in units of $\zeta/k=\tau_0$. } 
\begin{ruledtabular} 
\begin{tabular}{cccccccc}
$a$ & $N$ & $T_{\text{simulation}} \pm \delta T$  & $T_{\text{non-Markovian}}$  & $T_{\text{Markovian}}$ & $T_{\text{SSS}}$ & $T_{\text{RG}} $  & $T_{\text{AKH}}$ \\
\hline
$0.87$ 	& $50$ 	& $522 \pm 30$ 	& $543$	& $608$  &  $247$ & $491$		&$630$\\
 		& $100$ 	& $2040 \pm 60$ 	& $2096$ & $2340$ &  $707$ & $1962$		&$2220$\\
$4.33$   	& $50$   	& $128  \pm 2.5$ 	& $123$   & $246$  & $37.6$ & $491$		&$426$\\ 
   		& $100$ 	& $642  \pm 10$ 	& $599$  	& $1217$  &  $119$ &$1962$		&$1645$\\
   		& $300$ 	& $7439 \pm 350$ 	& $6933$ & $13583$ &  $1071$ &$17660$	&$14251$\\
\end{tabular} 
\end{ruledtabular} 
\end{table*}

In order to test the validity of the different theories, we compare it to the results of numerical simulations. For simplicity, we restrict ourselves to the case of the end-to-end cyclization reaction, where the reactive monomers are located at the extremities of the chain ($p=1,q=N$). For small values of $a$, we used directly the results of Brownian dynamic simulation with adaptative time step performed by Pastor \textit{et al.}\cite{Pastor1996}. We carried out supplementary simulations for larger values of $a$ and $N$ with the same algorithm and a smaller value of the time step. Apart from exploring supplementary ranges of parameters, these simulations enable us to analyze the statistics of the polymer reactive conformations, which has to our knowledge not been done before. 
In brief, the simulations use a Brownian dynamics algorithm and consist in generating stochastic trajectories that integrate the Langevin equation which corresponds to the Fokker-Planck equation (\ref{FKPRouseModes}). The fact that the time step is reduced when $\|\ve[R]\|-a$ becomes small is useful to increase the numerical precision. 

The  results for the reaction time are presented on Fig. \ref{FigureComparaisonSimuAvecSSS} and table \ref{TableauResultatsSimulationsTheories}, where the simulation results are compared to the various theories. 
As can be observed, the non-Markovian theory accurately predicts the values of the reaction time for all the values of $N$ and $a$. 
The non-Markovian result is almost always within the statistical error of the simulations,  
although the reaction times are estimated over large numbers of realizations which range from $2,500$ to $20,000$. 
In comparison, the results of the Markovian approximation are much less accurate, and the error made is roughly $100\%$ when the reactive radius $a$ is not too small. For small values of $a$ the Markovian approximation is however excellent, we will see below that the reason for this is that Markovian and non-Markovian theories predict the same asymptotic form of the reaction time $T$ in the limit of small $a$. The results of the harmonic spring approximation are less accurate than those of the Markovian theory, and can even differ by a factor  of $7$ from the cyclization time for the larger value of $N$. As mentioned above, the SSS theory predicts only the scaling relation $T\sim N^{3/2}/a$ for large $N$, contrarily to what is predicted in more precise theories and simulations. 
We also included in table \ref{TableauResultatsSimulationsTheories}  the results of the renormalization group theory calculated with Eq. (\ref{ResultRenormalizationGroup}), which does not predict correctly the values of the reaction time for $a=4.33$. Of course, the result of the renormalization group theory cannot be accurate in all regimes, as it is only asymptotically valid in the limit of large $N$. To be complete, we also included in table \ref{TableauResultatsSimulationsTheories} the recent results of Amitai \textit{et al.}\cite{Amitai2012}, who found that in the limit of small $a$, the reaction time is given by $T\simeq \sqrt{\pi/8}N^{3/2}/a+3A_3N^2$, where $A_3\simeq 0.04-0.055$ is a numerical coefficient obtained by fitting the data of numerical simulations. As can be seen in  table \ref{TableauResultatsSimulationsTheories}, this formula is not accurate for the larger values of $a$, which is not surprising since it is supposed to be valid only in the limit of very small values of $a$ ($a\ll1/\sqrt{N}$). 
In conclusion, the non-Markovian theory presented in this paper appears to be very accurate for all values of the parameters that we tried, to the difference of the existing theories that all have a limited range of validity. 

\section{The reactive shape of the polymer at cyclization and the asymptotic form of the reaction time on the non-Markovian theory}
\label{SectionReactiveShapeAndPositionMonomers}

\begin{figure}[h!]
\includegraphics[width=7.5cm,clip]{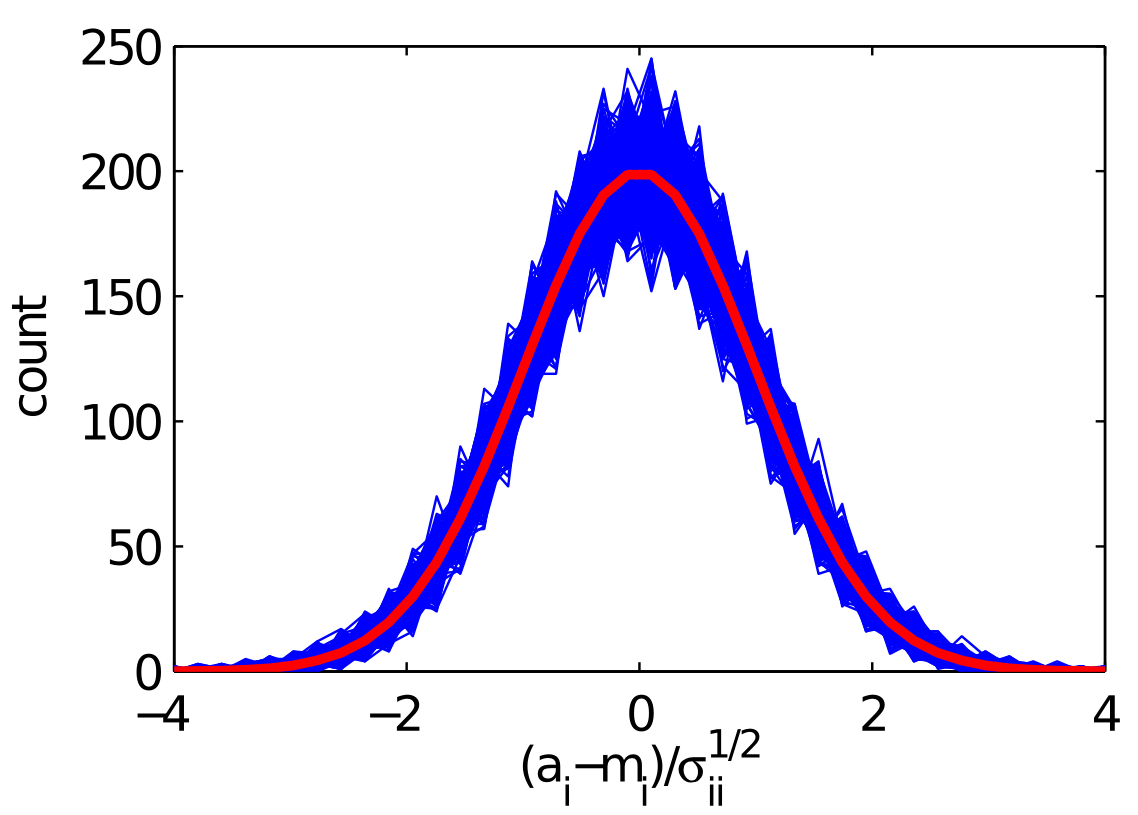}
\caption{Superposition of histograms of the modes $a_i$ in all spatial directions, for $N=300$ and $a=4.33$. For each value of $i$ ($2\le i\le N$), the values of $a_i$ at the instant of first cyclization are recorded in the simulation and rescaled by their average mean and variance over $2440$ realisations. The procedure is done for the radial and the two perpendicular spatial directions. All the resulting histograms appear in this figure, and are compared to the standard normal law (red thick curve). 
\label{FigHistogrammesModes}}
\end{figure}

\begin{figure}[h!]
\includegraphics[width=7.5cm,clip]{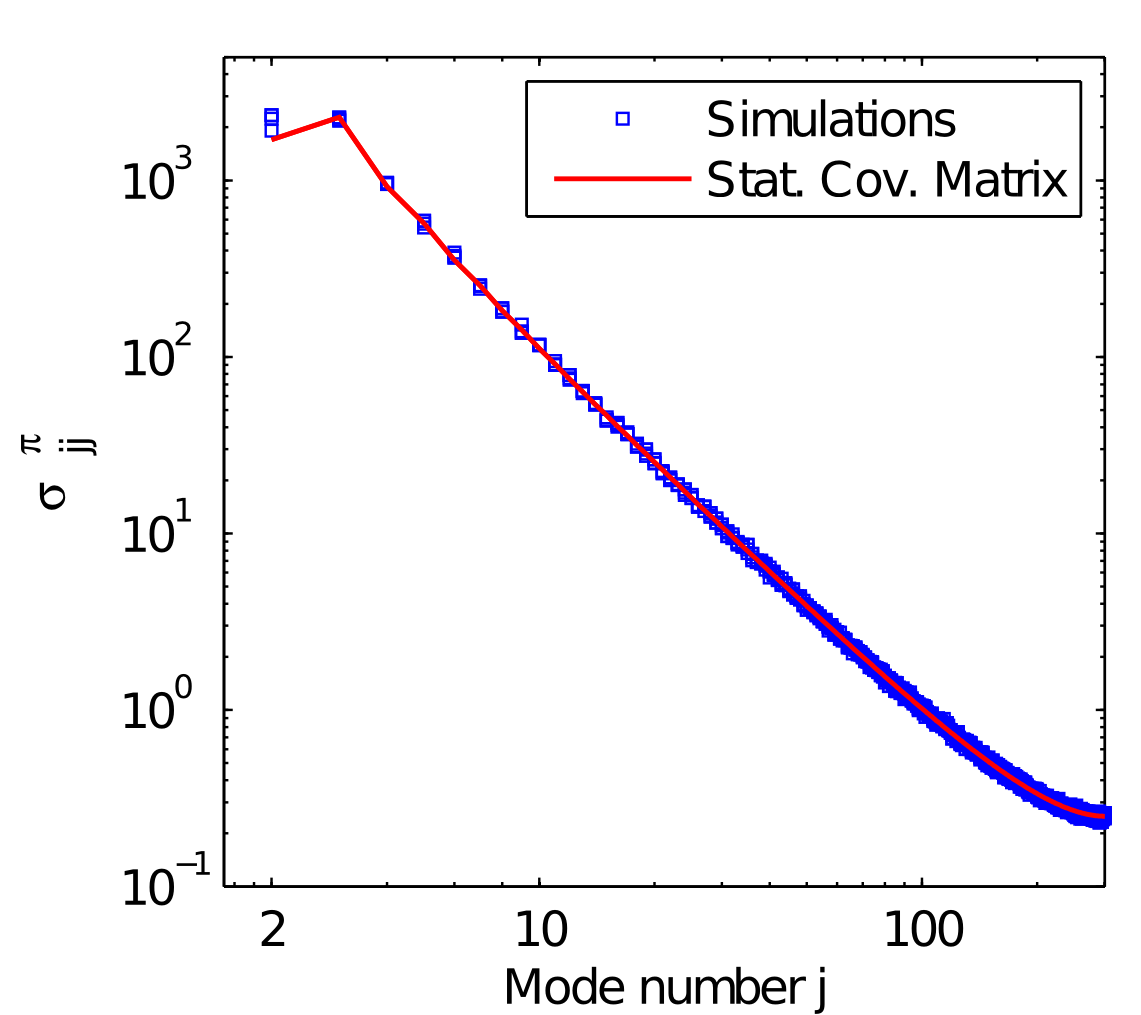}
\caption{Blue squares: diagonal terms $\sigma_{jj}^{\pi}$ ($2\le j\le N$) of the covariance matrix of the modes at the instant of first cyclization, in the 3 spatial directions. The red line represents the diagonal terms of the  covariance matrix of equilibrium looping conformations given by (\ref{StationaryCovarianceApprox}). Parameters: $N=300$ and $a=4.33$.  
\label{FigSIMADiag}}
\end{figure}

\begin{figure}[h!]
\includegraphics[width=7.5cm,clip]{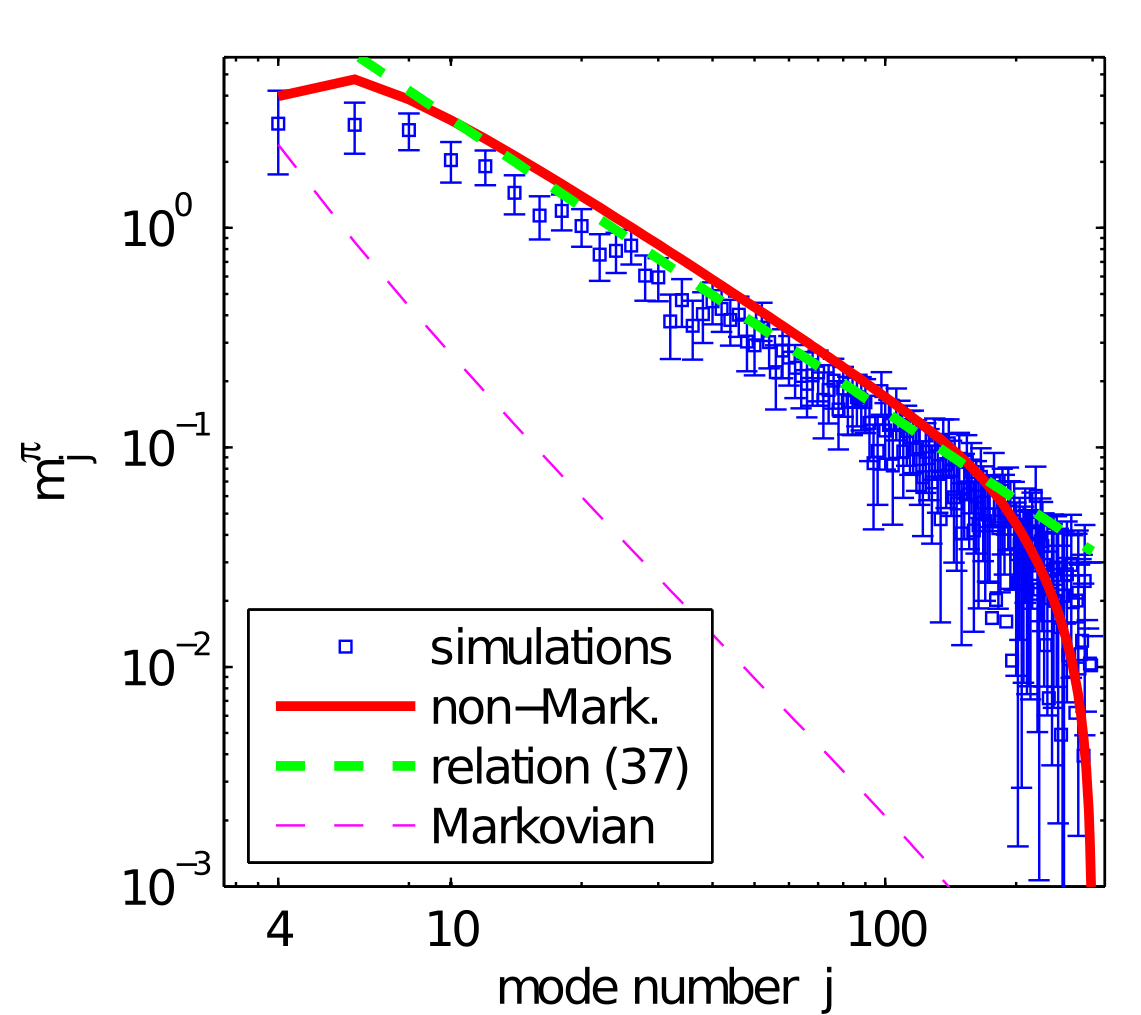}
\caption{Average value of the modes $m_j^{\pi}$ at the instant of end-to-end cyclization, for the same parameters as in figure \ref{ShapePolymerCyclization}. Only the modes for  $j$ even are shown (the other modes vanish). Symbols: simulations. Red line: non-Markovian theory. Green thick dashed line: asymptotic relation of the non-Markovian theory (\ref{scalingRelationMomentsCyclization}), proportional to $j^{-4/3}$. Magenta thin dashed line: Markovian approximation (\ref{Value_of_mq_Markovian}). The sign of the Markovian expression of $m_j^{\pi}$ has been inverted. Parameters: $N=300$ and $a=4.33$. 
\label{SpectrumPolymerCyclization}}
\end{figure}

\begin{figure}[h!]
\includegraphics[width=7.5cm,clip]{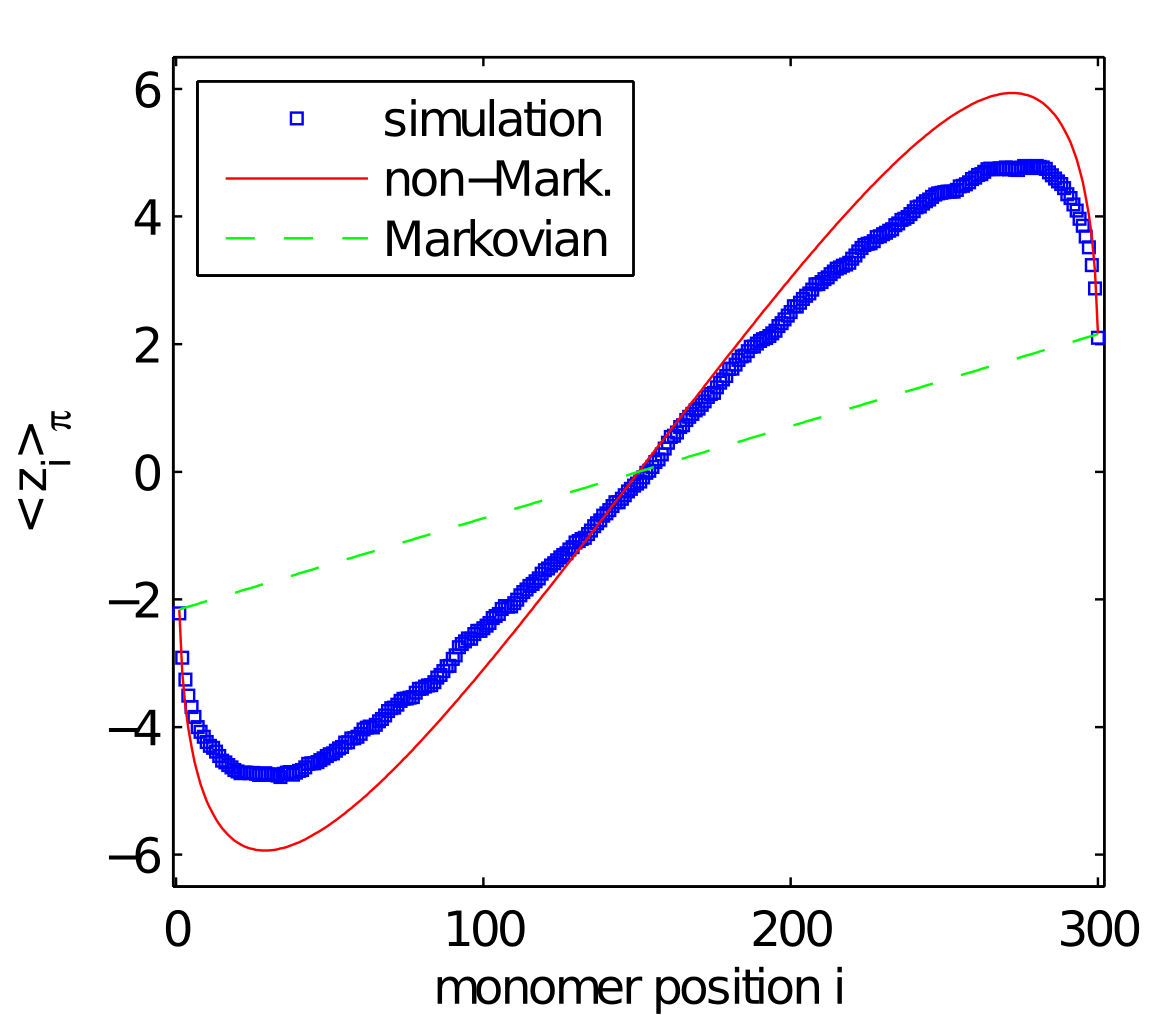}
\caption{Average position $\langle z_i\rangle $ of the $i^{th}$ monomer at the instant of first cyclization in the direction of reaction, as a function of the position of the monomer in the chain $i$. Parameters: $N=300$ and $a=4.33$. Blue symbols: results of simulations, averaged over $2440$ realizations. Red line: non-Markovian theory. Green dashed line: Markovian (Wilemski-Fixman) theory. Parameters: $N=300$ and $a=4.33$. \label{ShapePolymerCyclization}}
\end{figure}

As already mentioned, the key point that the classical Markovian theories do not consider is the out-of-equilibrium reactive polymer conformations. 
We now focus on the description of these reactive conformations in the non-Markovian theory and in the simulations, and  we determine whether or not the hypotheses of the non-Markovian theory are reasonable. We study in details the case $N=300$ and $a=4.33$, for which $2,440$ cyclization events have been recorded, and where the non-Markovian effects are important since the Markovian approximation overestimates the reaction time by a factor of $2$. The statistics of the reactive conformations is presented for these parameters in Figs. \ref{FigHistogrammesModes}, \ref{FigSIMADiag}, \ref{SpectrumPolymerCyclization}, \ref{ShapePolymerCyclization} and completes Fig. \ref{ExamplesConfigurationsCyclization}, where examples of reactive conformations in 3D are presented. 

The first hypothesis of the non-Markovian theory is that the splitting distribution $\pi$ is a multivariate Gaussian. This hypothesis is tested on Fig. \ref{FigHistogrammesModes}, where we represented the superposition of the histograms of all the modes $a_i$ in all spatial directions, after rescaling by the measured average and variance. If the Gaussian approximation were accurate, then all the histograms obtained in this way would look like a normal distribution, which is obviously the case in Fig. \ref{FigHistogrammesModes}. For each mode number $i$ and each spatial direction $\alpha$, the marginal probability density $\pi(a_{i,\alpha})$ looks like a normal distribution. Although we cannot readily deduce that the multivariate distribution $\pi(\vert \ve[a]\rangle)$ is Gaussian, these results indicate however that the Gaussian approximation is a very reasonable one.

Then, a second hypothesis of the non-Markovian theory is that the covariance matrix of the splitting distribution is approximated by the covariance matrix of equilibrium looping conformations given by Eq. (\ref{StationaryCovarianceApprox}). We represented the diagonal elements $\sigma_{jj}^{\pi}$ of this covariance matrix obtained in the theory and the simulations on Fig. \ref{FigHistogrammesModes}. As can be observed, the stationary covariance approximation is excellent for almost all modes whose number is larger than 3-4, but is not fully accurate for small mode numbers. 
We can therefore expect that the theory does not predict correctly the actual polymer conformation at the length scale associated to the first modes, which is of the order of $\sqrt{N}$. However, all the other diagonal terms $\sigma_{jj}^{\pi}$ are very well described by the stationary covariance approximation. We also investigated the values of the first non-diagonal terms $\sigma_{j,j+2}^{\pi}$, which are well described by their stationary value for small mode numbers. For larger values of the mode numbers, the noise of the simulated conformation (coming from statistical error and the finite value of the time step) is too large to even define the sign of the correlation coefficients $\sigma_{j,j+2}^{\pi}$ and to conclude whether the stationary covariance approximation is accurate for these correlations. 
Following these comments, we deduce that the stationary covariance approximation seems well supported by the comparison with simulations, although it could be supposed to fail to describe the polymer conformation at the large length scale $\sqrt{N}$. 

We now focus on the average shape of the polymer at the instant of reaction, which is the key quantity calculated in the non-Markovian theory. The average polymer shape is described by the average spectrum of the reactive conformations formed by the ensemble of the values $m_j^{\pi}$, which is represented on
Fig. \ref{SpectrumPolymerCyclization}. As can be observed, the predictions of the non-Markovian theory on the structure of the average spectrum are qualitatively correct for almost all values of the mode number $j$ and therefore all length scales. The non-Markovian theory slightly overestimates the values of $m_j^{\pi}$, especially for low values of the wave number $j$. This discrepancy between theory and simulations possibly comes from our simplifying hypotheses of a splitting distribution that is Gaussian with the stationary covariance approximation, which is not fully accurate for small mode numbers. 
However, the non-Markovian theory is much more precise than the Wilemski-Fixman Markovian theory, in which  the values of $m_j^{\pi}$ are given by their stationary value [Eq. (\ref{Value_of_mq_Markovian})]: the value of these modes have the wrong sign and are an underestimation of up to two orders of magnitudes  of the actual average value of the modes (Fig. \ref{SpectrumPolymerCyclization}). 
Coming back to the the space of monomer positions instead of modes, we can describe 
the average shape of the polymer by the function $\langle z_i\rangle_{\pi}=\sum_{j=2}^N Q_{ij}m_j^{\pi}$, which represents the average value of the spatial position of the $i^{\text{th}}$ monomer in the chain in the direction of reaction, and which is represented on Fig. \ref{ShapePolymerCyclization}. Importantly, the average shape of the polymer is found to be qualitatively  predicted by the non-Markovian theory (and not by the Markovian approximation), but this theory overestimates the magnitude of the polymer elongation at the instant of reaction. This failure of the theory at the large length scale $\sqrt{N}$ is eventually related to the stationary covariance approximation, which fails at these length scales.  The elongation of the polymer on average, as apparent on the average function $\langle z_i\rangle_{\pi}$ on Fig. \ref{ShapePolymerCyclization}, can also be seen directly on the pictures of the polymer reactive conformations, as shown in Fig. \ref{ExamplesConfigurationsCyclization}a and \ref{ExamplesConfigurationsCyclization}b, where we represented conformations issued from simulations and from the splitting distribution 
$\pi_{\Omega}$ of the non-Markovian theory. In these conformations, the polymer tends to be much more elongated in the direction of the reaction than in the equilibrium looping conformations shown on Fig. \ref{ExamplesConfigurationsCyclization}c. The fact that the polymer does not have to wait for reaching an equilibrium looping conformations so that the two end monomers come into contact implies a reaction kinetics that is faster than predicted by the Markovian Wilemski-Fixman theory. 

We now discuss some of the properties of the average spectrum $m_j^{\pi}$ of the reactive conformations. 
As can be seen on Fig. \ref{SpectrumPolymerCyclization}, the coefficients $m_j^{\pi}$ decrease as a power law of $j^{-\beta}$ over about one decade, with an exponent $\beta$ that is clearly less than 2. From Eq. (\ref{EquationFirstMomentDim3AveragedSimplified}), we can derive an analytical argument (presented in Appendix \ref{AppendixAsymptoticValueMoments}) that provides the asymptotic behavior of $m_j^{\pi}$ for large $j$, when the number of monomers tends to infinity while the rescaled reaction radius $\tilde{a}=a/\sqrt{N}$ is held constant:
\begin{equation}
	m_j^{\pi} \simeq \frac{ 2^{1/6} \ \tilde{a}^{1/3}}{\pi \ j^{4/3}} N \label{scalingRelationMomentsCyclization}.
\end{equation}
The value of $m_j^{\pi}$ predicted by this expression agrees reasonably well with the  structure of both the theoretical and simulated spectrums shown on Fig. \ref{SpectrumPolymerCyclization}. 
Such a slow decrease of the spectrum is transferred to the average shape $\langle z_i\rangle_{\pi}$, whose first derivative seems to be infinite at $i=1$ and $i=N$ on Fig. \ref{ShapePolymerCyclization}. 
This means that the first and last monomers of the chain are on average very shifted from the position of the reactive zone, thereby confirming the image of a polymer that forms an elongated loop at the instant of the reaction. When the size of the reactive region gets smaller ($\tilde{a}\rightarrow0$), (\ref{scalingRelationMomentsCyclization}) is not valid anymore, and is probably replaced by a law $m_j^{\pi}\sim N/j^{3/2}$, although our preliminary calculations suggest the existence of a logarithmic correction to this law. The difference between the slopes $3/2$ and $4/3$ is however difficult to detect in the simulations, where the data are noisy and the power-law behavior holds only over a decade.

In the limit $N\rightarrow\infty$, we can also get the qualitative form of the asymptotic behavior of the reaction time. 
We assume that $N\rightarrow\infty$ at fixed value of the rescaled reaction radius $\tilde{a}=a/\sqrt{N}$. The eigenvalues are well approximated by: $\lambda_j\simeq (j-1)^2\pi^2/N^2$, while the coefficients $b_j$ are given by: $b_j\simeq-\sqrt{8/N}$ if $j$ is even, and vanish for odd values of $j$. The time is rescaled by the power of $N$ that corresponds to the Rouse time: we pose $\tau=t/N^2$. 
Then, we find that the scaling $M_j=m_{j+1}^{\pi}/N$ is the correct scaling for the moments of the splitting distribution, as it leaves Equation. (\ref{EquationFirstMomentDim3AveragedSimplified}) invariant on $N$. We also introduce the rescaled reactive trajectory:
$Y_{\pi}(\tau)=R_{\pi}(\tau N^2)/\sqrt{N}$, and the rescaled mean square displacement $\Psi(\tau)= N\psi(t)$. Then, the reaction time is given in the limit of large $N$ by the relation:
\begin{align}
\frac{T}{N^2}=\int_0^{\infty}d\tau \frac{1}{(2\pi\Psi)^{3/2}} \left[\exp\left(-\frac{Y_{\pi}^2}{2\Psi}\right) - \frac{Z(\tilde{a},\Psi)}{Z(\tilde{a},1) } \right].
\end{align}
As the right-hand side term of this equation does not depend on $N$, we deduce the scaling relation $T\simeq N^2 F(\tilde{a})$ for large $N$. The function $F$ does not diverge for small values of its arguments, implying that for a small size of the reactive region, the reaction time reads $T\simeq N^2 F(0)$. This scaling relation is of the same type as the one that is deduced from the Markovian Wilemski-Fixman approach  (\ref{MarkovianScalingLargeN}), but the numerical coefficient $F(0)$ is different. The estimation of this numerical coefficient is rather difficult, because solving the equations for a small value of $a$ and a value of $N$ which is not large enough makes the result fall into the regime where the reaction time is $T\sim N^{3/2}/a$ and diverges with $a$. Our best estimation is $F(0)\simeq 0.175$, whereas the Markovian approximation gives $F(0)\simeq 0.2003$: the difference between the two numerical coefficients is about $12\%$. In intermediate regimes of $a$ and $N$, the results of the non-Markovian and Markovian theory differ by about $100\%$: in all cases, the non-Markovian effects, that result from the non-equilibrium reactive conformations of the polymer, give results that are quantitatively different from the Markovian approximation. 

Finally, we also consider the case of a small reactive radius: $a\rightarrow0$ at fixed $N$. Then, equation  (\ref{EquationFirstMomentDim3AveragedSimplified}) predicts that the moments $m_i^{\pi}$ are asymptotically proportional to $a$ in this limit, and that: $m_{i}^{\pi}=a b_i[-1+1/(\lambda_i L^2)]$. The fact that the moments are proportional to $a$ indicates that they are small and do not play any role in the reaction time. This can be seen by considering that $T$ in this limit is evaluated by taking the integrand of (\ref{EstimationMFPT_splitting_dim3_Explicite_Centered}) in the small time limit, where $R_{\pi}\simeq a(1+\mathcal{O}(t))$ has the same behavior as in the Markovian approximation. Therefore, the result of $T$ as $a\rightarrow0$ is exactly the time  (\ref{MarkovianScalingSmall_a}) predicted by the Markovian approximation, and it is also the same as predicted by the SSS theory. 
From the scaling argument (\ref{ScalingRelationQualitative}), the time $T\sim N^{3/2}/a$ comes from the diffusive regime of $\ve[R]$ at small times, where the monomers behave as if they were disconnected in an effective confining volume $N^{3/2}$. It is striking that, in this limit, all the approaches lead to the same result. It is likely to be linked to the fact that the determining step of the search process in this regime is diffusive, and that diffusion is a Markovian process. 

\section{Effect of the position of the reactive monomers on the reaction kinetics}
\label{SectionEffectPositionReactiveMonomers}
\begin{figure}[h!]
\includegraphics[width=7.5cm,clip]{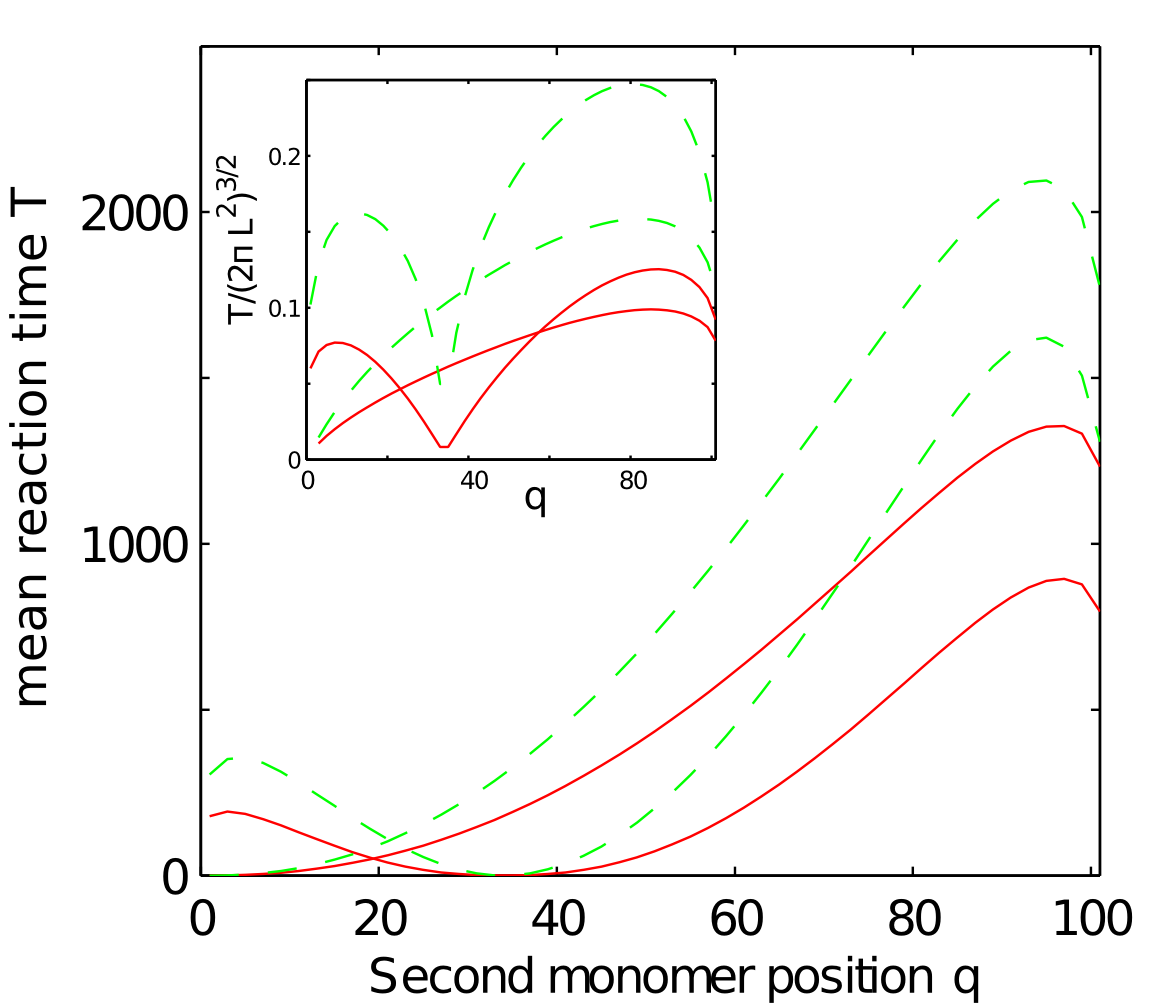}   
\caption{Variation of the reaction time with the position of the second reactive monomer $q$, when  the first monomer is at position $p=1$ (set of curves that vanish at $q=1$) and $p=34$ (curves that vanish at $q=34$). Continuous red lines: non-Markovian theory ; dashed green lines: Markovian theory. Parameters: $N=100$ and $a=2.01$. 
Inset: same figure after rescaling of the reaction time by the inverse of the $j$-factor $(2\pi L^2)^{3/2}$.}
\label{TempsReactionPositionMonomeresNonRescale}
\end{figure}

We now investigate the importance of the positions of the monomers on the reaction kinetics. We remind that $p$ and $q$ are the indexes of the two reactive monomers in the chain.
In Fig. \ref{TempsReactionPositionMonomeresNonRescale}, we show an example of how the reaction time varies with $p$ and $q$ for fixed values of $N$ and $a$. As expected, the reaction time vanishes when the reactive monomers are close ($p=q$), and the reaction time $T(p,q)$ increases in general when $\vert p-q\vert$ gets larger. However, 
in some regimes, the situation is more complex and $T(p,q)$ decreases when $\vert p-q\vert$ increases. This phenomenon occurs in particular when one of the monomers gets closer from the chain end  (Fig. \ref{TempsReactionPositionMonomeresNonRescale}), and possibly comes from the fact that the motion of an end-monomer in the subdiffusive regime is faster than for an interior monomer, whose motion is slowed down by the presence of two surrounding polymer chains. 
On the inset of Fig \ref{TempsReactionPositionMonomeresNonRescale}, we show the reaction time after it has been rescaled by the equilibrium contact probability density, which is also called the $j-$factor and reads: $j(p,q)=P_{\text{stat}}(\ve[R]=\ve[0])=1/(2\pi \vert p-q\vert)^{3/2}$. From this plot, it is quite obvious that the $j-$factor is not an accurate quantity to describe the dependance of the reaction time with the position of the reactants. 
This is not a surprise for diffusion controlled reactions and can be deduced from the expression of the reaction time (\ref{EstimationMFPT_splitting_dim3_Explicite_Centered}). This expression shows that the reaction time is inversely proportional to the $j-$factor, but that the other terms also contain a lot of information about the reaction time.

\begin{figure}[h!]
\includegraphics[width=7.5cm,clip]{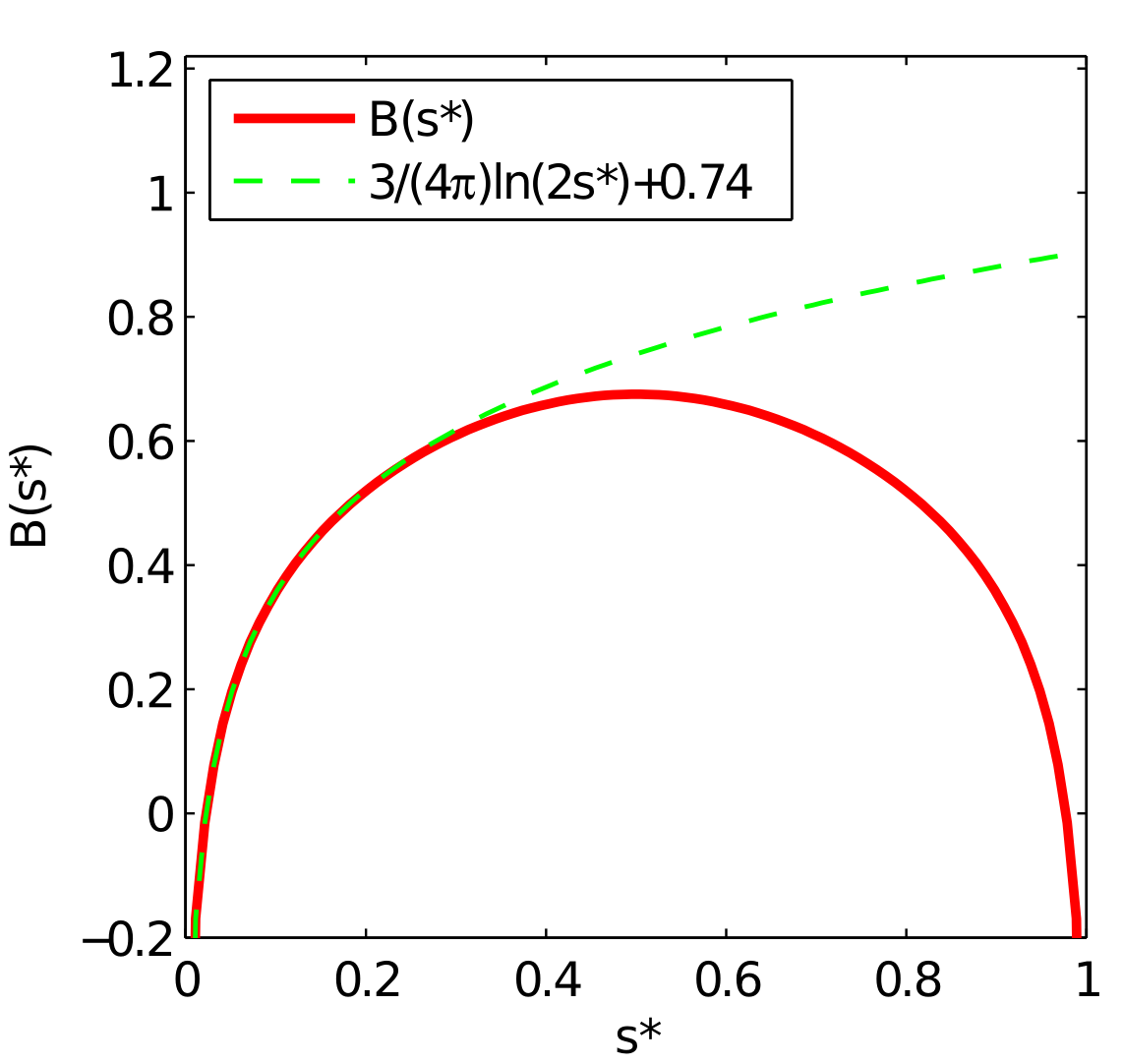}
\caption{The function $B(s^*)$ appearing in the expression (\ref{MFPT_InteriorMonomers}) of the reaction time for two close interior monomers, and its expansion for small values of $s^*$, given by (\ref{AsymptoticsBStar}). 
\label{FigFonctionB}}
\end{figure}

In order to illustrate this fact, we give asymptotic formulas for $T(p,q)$ in the Markovian approximation. 
First, in the limit $a\rightarrow0$ at fixed $p$ and $q$, we get the asymptotic result:
\begin{align}
	T\simeq \frac{\sqrt{\pi}\vert p-q\vert^{3/2}}{\sqrt{8}a} \ (a\rightarrow0)\label{RegimeSmallA},
\end{align}
which is the generalization of (\ref{MarkovianScalingSmall_a}) in the case of intramolecular reactions. 
In this case, the reaction time is effectively proportional to the inverse of the equilibrium contact probability density, with a proportionality factor that depends as $1/a$. In this regime, the reaction time comes from the diffusive behavior of the monomers at short times, it does not depend on the precise location of the monomers in the chain, but only on the number of monomers $\vert p-q\vert$ that separate them. 
When the number of monomers between the reactive groups grows to infinity however, the regime (\ref{RegimeSmallA}) disappears as the reaction time is controlled by the subdiffusive regime. A simple scaling argument enables to derive the scaling law for the reaction time in this regime. Considering that the vector $\ve[R]$ explores a volume of size $L\sim \vert p-q\vert^{1/2}$ with a subdiffusive walk of dimension $d_w=4$, we get that the reaction time scales as $T\sim L^{d_w}=\vert p-q\vert^2$. 
Actually, as shown below, this asymptotic form corresponds to the predictions of the Markovian theory,  
but with a numerical coefficient that depends on the position of the reactive monomers in the chain. 
We show in the appendix \ref{AppendixMarkovianEstimationDistanceReactiveMonomers} that the results of the Markovian theory must be discussed with the positions of the monomers. If one of the reactive monomers is an end-monomer (meaning that it is separated from the first monomer by a finite number of monomers as $N\rightarrow\infty$), then the reaction time reads in the regime $\vert p-q\vert\ll N$:
\begin{align}
	T\simeq c \ \vert p-q\vert^2 \ ; \ c \simeq 0.38 \label{MFPT_Interior_End_Monomers}
\end{align}
Note that the coefficient $c$ has an analytical form that is given in the appendix \ref{AppendixMarkovianEstimationDistanceReactiveMonomers}. If the two reactive monomers are interior monomers, then the reaction time depends on their average position on the chain $s^*$ defined by $s^*=(p+q)/(2N)$. We obtain the analytical formula: 
\begin{align}
	T\simeq \vert p-q\vert^2 \left[\frac{3}{4\pi} \ln \frac{N}{\vert p-q\vert} + B(s^*) \right]\label{MFPT_InteriorMonomers}
\end{align}
where $B(s^*)$ is a numerical function that can be determined explicitly (appendix \ref{AppendixMarkovianEstimationDistanceReactiveMonomers}) and that is represented on Fig. \ref{FigFonctionB}. 
This function describes how the reaction kinetics between two interior monomers is slowed down when the reactive monomers are located deeper in the chain. Interestingly,  (\ref{MFPT_InteriorMonomers}) indicates that there is a logarithmic correction to the scaling  $T\sim \vert p-q\vert^2$. 
Finally, we note that, for $s^*\rightarrow0$, $B(s^*)$ diverges as: 
\begin{equation}
	B(s^*)\simeq\frac{3}{4\pi}\ln(2s^*)+\kappa_0  \hspace{0.6cm} (s^*\rightarrow0\ ; \ \kappa_0\simeq0.74)\label{AsymptoticsBStar}.
\end{equation}
This means that the reaction time between two interior reactive groups that are located very close to the polymer extremity is approximately given by:
\begin{align}
	T\simeq \kappa_0 \vert p-q\vert^2 \ ; \ \kappa_0\simeq  0.74\label{Eq8391}
\end{align} 
We note that $\kappa_0\simeq 2 c$, where $c$ is the coefficient $c$  appearing in the expression (\ref{MFPT_Interior_End_Monomers}). This indicates that, even if $p\ll N$, the fact that the reactive monomers are inside the chain makes the reaction about twice slower than if the reactive monomers were separated by the same number of monomers, but with one of the reactive groups at an end-monomer. 
The expressions (\ref{MFPT_Interior_End_Monomers}),(\ref{MFPT_InteriorMonomers}),(\ref{Eq8391}) are asymptotically exact under the Markovian Wilemski-Fixman approximation, and the exact values of $c,\kappa_0$ and $B(s^*)$ are given in the appendix \ref{AppendixMarkovianEstimationDistanceReactiveMonomers}. We are not aware of any previous work where these expressions are derived, but we note that they are similar to those that are found with the renormalization group theory \cite{Friedman1993c}. Even if non-Markovian corrections are to be expected, these formulas illustrate the fact that the reaction time in the diffusion controlled regime is not well described by the $j$-factor, as it varies with the position of the reactants in the chain and as the scaling of $T$ is $\vert p-q\vert ^2 $ instead of $\vert p-q\vert^{3/2}$. 

\begin{figure}[h!]
\includegraphics[]{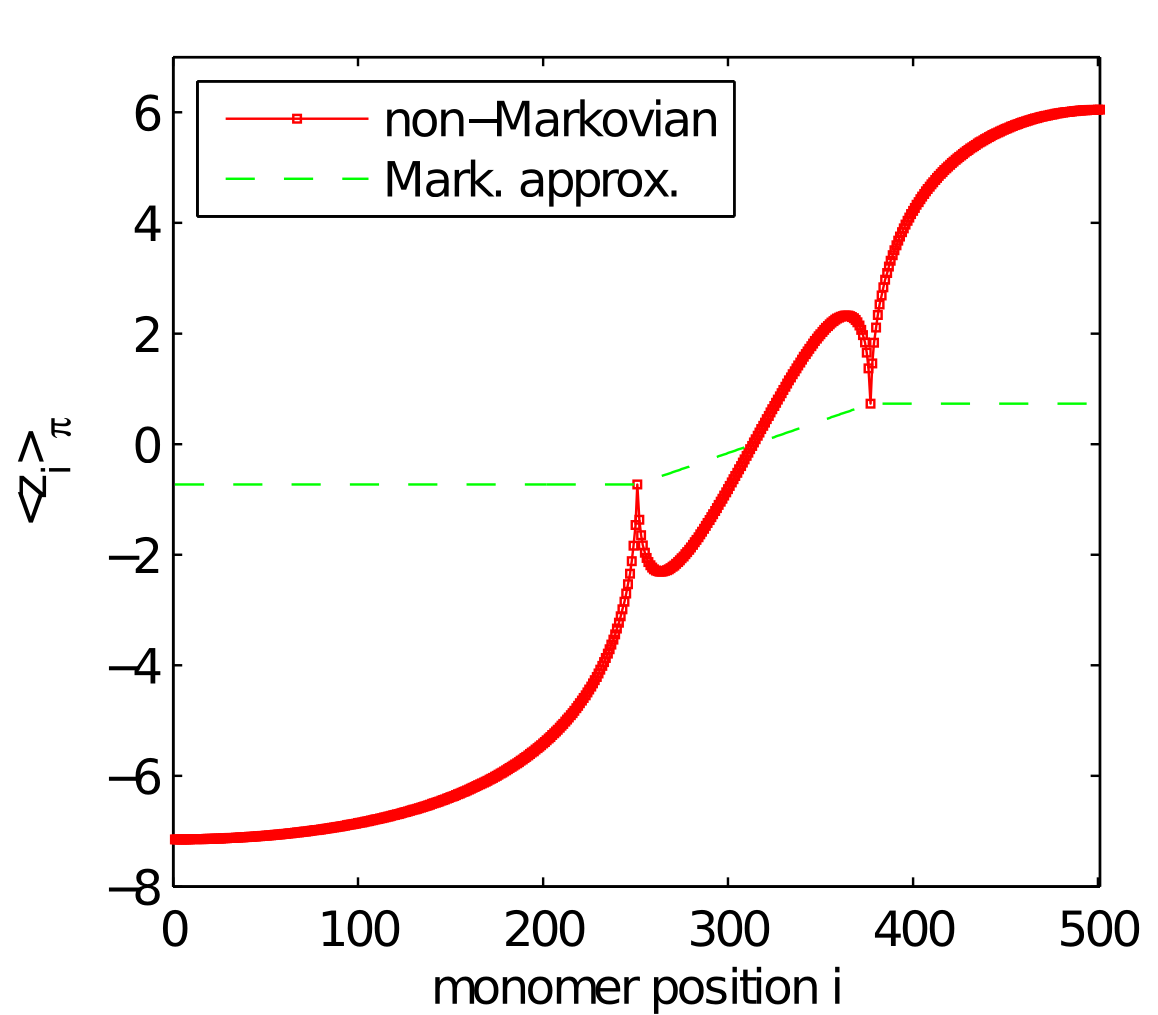}
\caption{Example of average positions of the monomers at the instant of the reaction in the case of an intramolecular reaction. There are $N=501$ monomers, and the reactive monomers are in positions $p=250$ and $q=376$. The  reactive radius is $a=1.46$.\label{FigExampleConformationInteriorMonomers}}
\end{figure}

Finally, we also show a typical shape of a reactive conformation when the reactive monomers are at the interior of the chain on Fig. \ref{FigExampleConformationInteriorMonomers}. We observe that the function $\langle z_i\rangle_{\pi}$ has singularities around the positions $i=p$ and $i=q$, and that the average reactive shape of the region of the polymer between 
the two reactive monomers is similar to the reactive conformations of a cyclizing polymer appearing on Fig. \ref{ShapePolymerCyclization}. Interestingly, the average shape of the parts at the exterior of the chain are also affected by the reaction: the whole polymer is therefore much more elongated in the direction of the reaction than in an equilibrium looping conformation, and this elongation is not limited to the part of the polymer between the reactive monomers. 

\section{Conclusion}
In conclusion,we have presented   in this paper  a non-Markovian theory that describes the kinetics of diffusion controlled intramolecular polymer reactions. This theory highlights the key role that is played by the conformational statistics of the polymer at the very instant of the reaction, which is not considered by classical theories of cyclization. For large chains, the reactive conformations are elongated in the direction of reaction and  are very different from equilibrium looping conformations. 
The average reactive conformations are characterized by a spectrum that decreases slowly with the wave number: the average reactive conformations present irregularities around the positions of the reactants in the chain, meaning that all the monomers that are next to the reactive monomers in the chain are at the instant of reaction very shifted in space from
the position they would have in an equilibrium looping configuration. In the case of end-to-end cyclization, the polymer forms an elongated loop at the instant of reaction. In the case of intramolecular reactions, the whole polymer is elongated, not only the part of the chain that lies between the reactants. The fact that the polymer does not need to wait to reach an equilibrium conformation where the reactants are brought into contact implies a reaction kinetics that is faster than the Wilemski-Fixman Markovian theory, which assumes that the reactive conformations are equilibrium looping ones. 

We have also reviewed the hypotheses of the existing theories of cyclization in the context of our more general non-Markovian theory. 
The SSS theory, which approximates the whole chain by a single spring, is exact for very small reactive radius but becomes rapidly inaccurate as $N$ becomes large, as it does not predict the correct scaling law $T\sim N^2$ for large $N$. The Wilemski-Fixman theory implicitly assumes that the reactive conformations are equilibrium looping conformations and relies therefore on a Markovian assumption. It predicts the two regimes $N^2$ and $N^{3/2}/a$ for the reaction time, but overestimates it quantitatively for large values of $N$. 
We have given a simple scaling argument that allows to understand the origin of these two regimes as the result of two-substeps in the reaction, which involve the diffusive and subdiffusive behavior of the reactants motion that appear at different time scales. Interestingly, the result of the renormalisation group theory can be recovered by taking the limit $d\rightarrow4$ of the Wilemski-Fixman expression of the reaction time for large $N$, suggesting that the renormalisation group approach also relies on a Markovian approximation. As a matter of fact, the difference between the result of the renormalization group and the Welemski-Fixman theory in the limit of large $N$ is only $2.4\%$. 

The non-Markovian theory presented in this paper predicts the same scaling laws as the Markovian theory, but with a different numerical coefficient in the large $N$ regime, and is found to be in very good agreement with simulations for all the values of parameters that we tried. This demonstrates that the fact of taking into account the reactive conformations is the essential missing ingredient of existing Markovian theories. The non-Markovian theory still makes approximations, as it assumes the Gaussianity of the distribution of reactive conformations with a stationary covariance matrix. These hypotheses are well supported by the comparison with simulations: even if the non-Markovian theory leads to an overestimation of the average value of the modes at the instant of reaction, it catches the correct structure of the spectrum of the reactive conformations at all length scales. 
The fact that the Markovian and the non-Markovian theory predict the same regimes for the reaction time is not obvious. Actually, it is not always the case: 
in the case of intermolecular reactions in a 1D space, the Markovian approximation has recently been shown to lead to incorrect scaling laws that can overestimate the true reaction time by several orders of magnitudes\cite{guerin2012b}. In the different context of the study of other stochastic processes such as Fractional Brownian Motion, it has also been shown that a Markovian approach can give wrong scaling exponents that characterize the first passage time density\cite{Sanders2012}. 

Finally, we also gave formulas that show how the reaction time varies with the positions of the reactants in the chain  in the Markovian approximation. These formulas highlight the fact that the reaction time cannot be deduced by considering only the contact probability density of the two reactants. In  the regime of a continuous chain, for a fixed distance between the reactants, the reaction time depends on the location of the reactants in the chain, and is slowed down when the reactive monomers are in the interior of the chain rather than in the exterior. 

In this paper, we only considered the simple Rouse chain model of polymer, for which the reaction kinetics is already non-trivial to determine. In future works, we aim at estimating the non-Markovian effects appearing in reactions involving complex polymers or general non-Markovian processes. 

\section*{Acknowledgements}
Support from European Research Council starting Grant FPTOpt-277998 and the French National Research Agency (ANR) Grants Micemico and DynRec are acknowledged.


\appendix

\section{Some useful properties of Gaussian processes: projection and propagation formulas}
\label{AppendixProjectionPropagationFormulas}
We remind here some properties related to the Gaussian processes. First, we remind formulas for conditional probabilities. Let us consider a Gaussian distribution of $N$ variables ($a_1,...a_N$), with covariance matrix $\sigma_{ij}$ and mean vector $m_i$. Consider also the variable $X=\sum_{i=1}^{N} b_i a_i=\langle b\vert a\rangle$, which is a linear combination of the original variables $a_i$. Then, the probability density distribution of $a_i$, given that $X$ takes a particular value $X_0$, is a Gaussian. We note $m_i^X$ its average and $\sigma_{ij}^*$ the covariance matrix of the conditional distribution, given by:
\begin{align}
	&m_i^{X}=m_i-\frac{\langle e_i\vert \sigma\vert b\rangle }{\langle b\vert\sigma\vert b\rangle}(\langle b\vert m\rangle-X)\label{Projection1},\\
	&\sigma_{ij}^*=\sigma_{ij}-\frac{\langle e_i\vert \sigma\vert b\rangle\langle e_j\vert \sigma\vert b\rangle}{\langle b\vert \sigma\vert b\rangle}\label{Projection2},
\end{align}
where $\vert e_i\rangle$ represents the basis vector whose all components vanish except the one in $i^{\text{th}}$ position, whose value is $1$.  
(\ref{Projection1}) and (\ref{Projection2}) are called projection formulas. As explained in details in a previous work\cite{guerin2012b}, these formulas can  be derived from the formulas on conditional distributions presented in the book of Eaton\cite{Eaton1983}.

Then, we also remind how a Gaussian distribution is propagated under one dimensional version of the Fokker-Planck equation (\ref{FKPRouseModes}), which reads:
\begin{align}
\frac{\partial P}{\partial t}=\lambda_i  \sum_{i=1}^{N} \frac{\partial}{\partial a_i}  (a_i P)+ \sum_{i=1}^{N}\frac{\partial^2P}{\partial a_i^2}.\label{FKPRouseModesAppendix}
\end{align}
Let us assume that at $t=0$, the distribution $P(a_1,...,a_N,t=0)$ is a Gaussian, with mean vector $m_i$ and covariance matrix $\sigma_{ij}$. Then, if $P$ satisfies the Fokker-Planck equation (\ref{FKPRouseModesAppendix}), it remains Gaussian at all later times $t>0$ with a mean vector $\mu_i(t)$ and a covariance matrix $\gamma_{ij}(t)$ which read\cite{VanKampen1992}:
\begin{align}
&\mu_i(t)=m_i e^{-\lambda_i t}\label{Propagation1},\\
&\gamma_{ij}(t)=\sigma_{ij} e^{-\lambda_i t}e^{-\lambda_j t} +\frac{\delta_{ij}}{\lambda_i}(1-e^{-2\lambda_i t})\label{Propagation2}.
\end{align}
The two formulas (\ref{Propagation1}) and (\ref{Propagation2}) are called propagation formulas. 

\section{The functions $\phi(t)$ and $\psi(t)$}
\label{AppendixFunctionPhiAndPsi}
Here, we describe how to derive the effective propagator $P(\ve[R],t\vert\{\ve[R]_0,\text{stat}\})$. Because all coordinates are independent, we consider the same problem for the first coordinate $X$ of $\ve[R]$. Then, the modes can be considered as scalars, and our aim is to calculate $P(X,t\vert \{X_0,\text{stat}\})$ the probability to observe $X$ at $t$ given that initially the polymer is at equilibrium with an initial position $X_0$. 
Using the projection formula (\ref{Projection1}), we find that the average $m_i^{\text{stat},X_0}$ of the mode $a_{i,x}$ at equilibrium given that $X=X_0$ writes:
\begin{align}
	m_i^{\text{stat},X_0}=X_0\frac{\langle e_i \vert \sigma^{\text{stat}}\vert b\rangle}{\langle b \vert \sigma^{\text{stat}}\vert b\rangle} = \frac{X_0 b_i}{L^2\lambda_i}.
\end{align}
Then, using the propagation formula (\ref{Propagation1}), we can compute the average of $X$ at $t$, for the same initial conditions:
\begin{align}
	\mathbb{E}(X,t\vert  \{ X_0,\text{stat}\},0) &= \sum_{i=2}^N b_i m_i^{\text{stat},X_0} e^{-\lambda_i t} \nonumber\\
	&=X_0 \sum_{i=2}^N \frac{b_i^2 e^{-\lambda_i t}} {L^2\lambda_i}=X_0\phi(t)\label{CalculDePhi}.
\end{align}
From this equation, we easily deduce the value (\ref{DefinitionPhi}) of the function $\phi(t)$ given in the main text. 

The variance is computed the same way. We call $\sigma_{ij}^{\text{stat}}$ the covariance matrix of the modes $a_i,a_j$ at equilibrium. According to the projection formula (\ref{Projection2}), the covariance of the modes $a_i$, $a_j$ at equilibrium with the condition that  $X=X_0$ is:
\begin{align}
\sigma_{ij}^{\{\text{stat},X_0\}}=\sigma^{\text{stat}}_{ij}-\frac{\langle e_i \vert \sigma^{\text{stat}}\vert b\rangle \langle e_j \vert \sigma^{\text{stat}}\vert b\rangle }{\langle b \vert \sigma^{\text{stat}}\vert b\rangle },
\end{align}
Using the equilibrium value $\sigma_{ij}^{\text{stat}}=\delta_{ij}/\lambda_i$, we find:
\begin{align}
\sigma_{ij}^{\{\text{stat},X_0\}}=\frac{\delta_{ij}}{\lambda_i}-\frac{b_i b_j }{\lambda_i \lambda_j L^2}\label{DefinitionStationaryCovarianceMatrix}.
\end{align} 
We now define $\gamma_{ij}^{\{\text{stat},X_0\}}$ the covariance of the modes $i,j$ at $t$, when the polymer is initially at equilibrium with the condition $X=X_0$. Using the propagation formula (\ref{Propagation2}), we get:
\begin{align}
	\gamma_{ij}^{\{\text{stat},X_0\}}&=\sigma_{ij}^{\{\text{stat},X_0\}}e^{-(\lambda_i+\lambda_j) t} +\frac{\delta_{ij}}{\lambda_i} (1-e^{-2\lambda_j t})\nonumber\\
	&=\frac{\delta_{ij}}{\lambda_i}-\frac{b_i e^{-\lambda_i t} b_j e^{-\lambda_j t}}{\lambda_i\lambda_j L^2}=\gamma_{ij}^{\{\text{stat},*\}}\label{Eq723}.
\end{align}
The last line states that $\gamma_{ij}^{\{\text{stat},X_0\}}$ can also be noted $\gamma_{ij}^{\{\text{stat},*\}}$, as it does not depend on $X_0$. Summing the expression (\ref{Eq723}) over $i,j$ (after multiplication by $b_i b_j$) gives the expression of $\psi(t)$, which is the variance of $X$ at $t$, for equilibrium initial conditions  with $X=X_0$:
\begin{align}
	\psi(t)&=\sum_{i,j=2}^{N} b_i b_j \gamma^{\{\text{stat},X_0\}}_{ij}\nonumber \\
	&=\sum_{i=2}^N \frac{b_i^2}{\lambda_i}-\sum_{i,j=2}^N \frac{b_i^2 e^{-\lambda_i t} b_j^2 e^{-\lambda_j t}}{\lambda_i\lambda_j L^2}.
\end{align}
Using the expression (\ref{ValueOfEquilibriumLength}) for $L$, and comparing with the value (\ref{CalculDePhi}) of $\phi(t)$, we find that:
\begin{align}
	\psi(t)=L^2[1-\phi(t)^2]\label{RelationPhiPsi}.
\end{align}
This equation is the value (\ref{DefinitionPhi}) of $\psi$ given in the main text. 
Coming back to the 3D case, because $\ve[R]$ is a Gaussian process, it is totally defined by its variance and average, and the effective propagator is:
\begin{align}
P(\ve[R],t\vert \{\text{stat},\ve[R]_0\},0)=\frac{1}{(2\pi \psi)^{3/2}}\exp\left\{-\frac{(\ve[R]-\phi \ve[R]_0)^2}{2\psi}\right\}.
\end{align}
We also give the behavior of $\phi$ in the limit $N\rightarrow\infty$. We pose $\tau=t/N^2$ and $\Phi(\tau)=\phi(t/N^2)$. We also introduce $s_p=(p-1/2)/N$ and $s_q=(q-1/2)/N$, which are the coordinate of the reactive monomers in the chain. Then, the rescaled function $\Phi$ reads:
\begin{align}
	\Phi(\tau)=\frac{2N }{L^2}\sum_{j=1}^{\infty}[\cos(s_p \pi j)-\cos(s_q \pi j)]^2 \frac{e^{-j^2\pi^2\tau}}{j^2\pi^2}.
\end{align}
Note that $L^2=\vert p-q\vert = N \vert s_q-s_q\vert$ 
If $\tau\rightarrow0$, the series can be transformed into an integral: we pose $y=j\sqrt{\tau}$ and we obtain:
\begin{align}
	1-\Phi(\tau)\simeq \frac{2 N\alpha }{L^2}\int_{0}^{\infty}\frac{dy}{\sqrt{\tau}} \frac{1-e^{-y^2\pi^2}}{y^2 \pi^2 \tau^{-1}}
	=\frac{2 N\alpha \sqrt{\tau}}{\sqrt{\pi}L^2}\label{AnomalousBehaviorPhi},
\end{align}
where $\alpha$ is the average value (over $j$) of the slowly varying term $[\cos(s_p \pi q)-\cos(s_q \pi j)]^2$. In the case of two monomers at the interior of the chain ($0<s_p<1$ and $0<s_q<1$), we have $\alpha=1$. In the case of the cyclization, where the two reactive monomers are at the chain extremities, we obtain $\alpha=2$. In the case of the reaction between an end-monomer and an interior-monomer, $\alpha=3/2$. Finally, the anomalous behavior (\ref{AnomalousBehaviorPhi}) of $\Phi$ is transferred to the function $\psi$, which is equal to:
\begin{align}
	\psi(t)\simeq \kappa\sqrt{t} \ ; \ \kappa=4\alpha  / \sqrt{\pi} \ ; 1\ll t\ll N^2\label{ValueOfkappa}.
\end{align}

\section{Derivation of the self-consistent equations for the moments $m_i^{\pi}$ in the non-Markovian theory.}
\label{AppendixDerivationEquationMPi}

In this section, we derive the equation (\ref{EquationFirstMomentDim3AveragedSimplified}) in the main text. We remind that $\Omega$ represents the direction of the vector $\ve[R]$ at the instant of reaction, and that $\pi_{\Omega}(\vert\ve[a]\rangle)$ is the distribution of variables $\ve[a]_i$ at the instant of reaction, given that the direction of $\ve[R]$ is $\Omega$. We note $\ve[u]_r$ the unit vector pointing in the direction $\Omega$, while $\theta$ and $\varphi$ refer to the polar and azimuthal angles. 
The key hypothesis of the non-Markovian theory is that $\pi_{\Omega}$ is a multivariate Gaussian. For symmetry reasons, the average vector $\ve[a]_i$ at the instant of reaction points in the direction $\Omega$ and is given by  $m_i^{\pi}\ve[u]_r(\Omega)$. We also make the simplifying assumption that the covariance matrix of the splitting distribution is given by the stationary covariance matrix (conditioned to a particular value of $\ve[R]$, so that it is given by Eq. (\ref{DefinitionStationaryCovarianceMatrix}), and the only unknown variables of the theory are the average moments $m_i^{\pi}$ and the reaction time $T$. 
In this appendix, we first derive the equations in the case that the initial distance between the reactants is fixed to a single value $R_0$, and at the end we will show how to adapt the calculation to a distribution of initial distances. 
We start from an integral equation that is deduced from (\ref{EquationIntegraleSplitting}) after multiplication by $\delta(\langle b\vert \ve[a]\rangle-\ve[R]_f)$, and that defines both $T$ and $\pi_{\Omega}$:
\begin{align}
&	T P_{\text{stat}}(\ve[R]_f)P_{\text{stat}}(\vert \ve[a]\rangle\vert \ve[R]_f)= \nonumber\\
		&\int_0^{\infty}dt \int d\Omega [ P(\ve[R]_f,t\vert \pi_{\Omega},0)P(\vert \ve[a]\rangle,t\vert \ve[R]_f,t;\pi_{\Omega},0)\nonumber\\
&-P(\ve[R]_f,t\vert \{\text{stat},R_0\ve[u]_r\},0)P(\vert \ve[a]\rangle,t\vert \ve[R]_f,t;\{\text{stat},R_0\ve[u]_r\},0)]\label{EQB11}.		
\end{align}
Integrating (\ref{EQB11}) over all conformations, and noting that the conditional distributions 
$P(\vert \ve[a]\rangle,t\vert \ve[R]_f,t;...)$ are normalized, we get the following expression for the reaction time:
\begin{align}
	&T P_{\text{stat}}(\ve[R]_f)=\nonumber\\
	&\int_0^{\infty}dt \int d\Omega [ P(\ve[R]_f,t\vert \pi_{\Omega},0)
-P(\ve[R]_f,t\vert \{\text{stat},R_0\ve[u]_r\},0)]\label{valueMeanReactionTimeAppendix}.
\end{align}
We can assume without loss of generality that $\ve[R]_f=R_f \ve[u]_z$, with $\ve[u]_z$ the unit vector pointing in a fixed (arbitrary) direction. 
Self-consistent equations for the moments $m_i^{\pi}$ will be derived by multiplying (\ref{EQB11}) by $a_{iz}$ and  integrating over the conformations. Some intermediate calculations need to be done. First, we calculate the following integral, which is interpreted as the expression of an average quantity over a conditional distribution, and is readily deduced from the projection formula (\ref{Projection1}):
\begin{align}
	\int d\vert \ve[a]\rangle a_{iz} P_{\text{stat}}(\vert \ve[a]\rangle\vert R_f \ve[u]_z) =m_i^{\{\text{stat},R_f\}}= \frac{R_f b_i}{\lambda_i L^2}\label{Eq783}.
\end{align}
Then, using (\ref{Propagation1}), we note that the average value of $a_i$ at $t$ in the radial direction $\ve[u]_r$, given that the initial distribution is the splitting probability $\pi_{\Omega}(\vert \ve[a]\rangle)$, is given by: $\mu_i(t)=m_i^{\pi}e^{-\lambda_i t}$. Because we assume that the covariance matrix of $\pi_{\Omega}$ is the stationary covariance matrix, the covariance of $a_i,a_j$ at $t$ is equal to $\gamma_{ij}^{\text{stat},*}$. Hence, using the projection formula (\ref{Projection1}), we can calculate the following integral (at fixed value of the angle $\theta$): 
\begin{align}	
\int d\vert\ve[a]&\rangle a_{ir} P(\vert \ve[a]\rangle ,t\vert R_f \ve[u]_z,t ; \pi_{\Omega},0) =\nonumber\\  &m_i^{\pi} e^{-\lambda_i t}- \frac{\langle e_i\vert \gamma^{\text{stat},*} \vert b\rangle}{\psi} (\langle b\vert \mu\rangle - R_f\cos\theta)\label{EqC13},
\end{align}
where we used the fact that the projection of $R_f \ve[u]_z $ over the direction $\ve[u]_r$ is $R_f \cos\theta \ve[u]_r$. We introduce the reactive trajectory $R_{\pi}(t)$ in the direction of the reaction:
\begin{align}
R_{\pi}(t)=\sum_{i=2}^N b_i \mu_i=\sum_{i=2}^N b_i m_i e^{-\lambda_i t}\label{Eq435}.
\end{align}
Using (\ref{Eq723}), we get the simplification:
\begin{align}
\langle e_i \vert \gamma^{\{\text{stat},*\}}\vert b\rangle=\frac{b_i(1-\phi e^{-\lambda_i t})}{\lambda_i}\label{4652},
\end{align}
and we also introduce $\mu_i^{\pi,0}$, which is interpeted as the average value of the mode $a_i$ in the radial direction, conditioned to the fact that $\ve[R]=0$:
\begin{align}
\mu_i^{\pi,0}=m_i^{\pi}e^{-\lambda_i t} - \frac{R_{\pi} b_i(1-\phi(t)e^{-\lambda_i t})}{\lambda_i\psi(t)}\label{1938}.
\end{align}
From (\ref{Eq435}),(\ref{4652}),(\ref{1938}) we find that (\ref{EqC13}) can be rewritten as:
\begin{align}	
\int d\vert\ve[a]\rangle a_{ir} P(\vert \ve[a]\rangle ,t&\vert R_f \ve[u]_z,t ; \pi_{\Omega},0) =\nonumber\\  &\mu_i^{\pi,0} + \frac{ b_i (1-\phi\ e^{-\lambda_i t})}{\lambda_i \psi} R_f\cos\theta\label{Eq988}.
\end{align}
With the same method, we calculate the following integral:
\begin{align}	
\int d\vert\ve[a]\rangle a_{i\theta} P(\vert \ve[a]\rangle ,t\vert &R_f \ve[u]_z,t ; \pi_{\Omega},0) = \nonumber\\
& - \frac{ b_i (1-\phi\ e^{-\lambda_i t})}{\lambda_i \psi} R_f\sin\theta\label{Eq989}.
\end{align}
Now, noting that $a_{iz}=a_{ir} \text{cos}\theta-a_{i\theta}\text{sin}\theta$ and using the two evaluations (\ref{Eq988}),(\ref{Eq989}), we get:
\begin{align}	
&\int d\vert\ve[a]\rangle a_{iz}P(\vert \ve[a]\rangle ,t\vert R_f \ve[u]_z,t ; \pi_{\Omega},0)\nonumber\\
&=\cos\theta m_i^{\pi} e^{-\lambda_i t} - \frac{ b_i (1-\phi\ e^{-\lambda_i t})}{\lambda_i \psi} (\cos\theta R_{\pi}-R_f) \label{Eq835}.
\end{align}
With the same reasoning, we get that:
\begin{align}	
&\int d\vert\ve[a]\rangle a_{iz}P(\vert \ve[a]\rangle ,t\vert R_f \ve[u]_z,t ; \{R_0 \ve[u]_r,\text{stat} \},0)=\nonumber\\
&\cos\theta \frac{b_i R_0}{\lambda_i L^2} e^{-\lambda_i t} - \frac{ b_i (1-\phi\ e^{-\lambda_i t})}{\lambda_i \psi} (\cos\theta R_0\phi-R_f).
\end{align}
Using the relation (\ref{RelationPhiPsi}), we simplify this result:
\begin{align}	
\int d\vert\ve[a]&\rangle a_{iz}P(\vert \ve[a]\rangle ,t\vert R_f \ve[u]_z,t ; \{R_0 \ve[u]_r,\text{stat} \},0)=\nonumber\\
&\cos\theta \frac{b_i R_0}{\lambda_i \psi} (e^{-\lambda_i t}-\phi) + \frac{ b_i (1-\phi\ e^{-\lambda_i t})}{\lambda_i \psi} R_f\label{Eq836}.
\end{align}
Now, the propagator $P(R_f \ve[u]_z,t\vert \pi_{\Omega},0)$ is given by:
\begin{align}
	P(R_f \ve[u]_z,t\vert \pi_{\Omega},0) =\frac{1}{[2\pi\psi]^{3/2}}\ \exp \left\{-\frac{(R_f \ve[u]_z -R_{\pi}\ve[u]_r)^2}{2\psi }\right\}.	
\end{align}
We expand this expression at first oder in $R_f$:
\begin{align}
	P(R_f \ve[u]_z,t\vert \pi_{\Omega},0) \simeq \frac{e^{-R_{\pi}^2/(2\psi)}}{[2\pi\psi]^{3/2}}\left(1+\frac{R_{\pi}R_f}{\psi}\cos\theta+\mathcal{O}\label{Eq899}(R_f^2)\right).	
\end{align}
We also get the small $R_f$ expression of the second propagator $P(R_f \ve[u]_z,t\vert \{\text{stat},R_0\ve[u]_r\},0)$:
\begin{align}
	P(R_f \ve[u]_z,t&\vert \{\text{stat},R_0\ve[u]_r\},0) \simeq \nonumber \\
	&\frac{e^{-(R_0\phi)^2/(2\psi)}}{[2\pi\psi]^{3/2}}\left(1+\frac{R_0\phi R_f}{\psi}\cos\theta+\mathcal{O}(R_f^2)\right)	\label{Eq890}.
\end{align}
Using equations (\ref{Eq783}),(\ref{Eq835}),(\ref{Eq836}),(\ref{Eq899}),(\ref{Eq890}), we multiply (\ref{EQB11}) by $a_{iz}$, integrate it over the conformations and develop the result in powers of $R_f$. The term proportional to $(R_f)^0$ will vanish after integration over the angle $\theta$, so that we write the term proportional to $R_f$, which reads:
\begin{align}
&T P_{\text{stat}}(\ve[0])\frac{b_i}{\lambda_i L^2}= \nonumber\\
&\int_0^{\infty}dt \int_0^{\pi}\frac{d\theta \sin\theta }{2}  \frac{1}{[2\pi\psi]^{3/2}}  \times \nonumber\\
\Bigg\{&e^{-\frac{R_{\pi}^2}{2\psi}}
\left[\frac{R_{\pi}}{\psi} \mu_i^{\pi,0} (\cos\theta)^2 +\frac{b_i(1-\phi e^{-\lambda_i t})}{\lambda_i \psi}\right]\nonumber\\
&-e^{-\frac{(R_0\phi)^2}{2\psi}}\left[\frac{b_i R_0^2\phi}{\lambda_i \psi^2} (e^{-\lambda_i t}-\phi)(\cos\theta)^2+\frac{b_i(1-\phi e^{-\lambda_i t})}{\lambda_i \psi}\right]\Bigg\}.
\end{align}
Finally, performing the integration over $\theta$, and taking account the relation (\ref{valueMeanReactionTimeAppendix}) (written for $\ve[R]_f=\ve[0]$), we find the simplified form of the self-consistent equations that define the non-Markovian theory:
\begin{align}
&\int_0^{\infty}\frac{dt}{\psi^{5/2}} \Bigg\{ \left(\frac{\mu_{i}^{\pi,0}R_{\pi}}{3}+ 
\frac{b_i\phi (\phi- e^{-\lambda_i t})}{\lambda_i } \right)\text{exp}\left(-\frac{R_{\pi}^2}{2\psi}\right) \nonumber\\
&- \frac{b_i \phi (\phi-e^{-\lambda_i t}) }{\lambda_i}\left(1-\frac{R_0^2}{3\psi}
\right) \text{exp}\left(-\frac{\phi^2 R_0^2}{2\psi}\right)\Bigg\}=0, 	
\end{align}
which is the form of the self-consistent equations that define the first moments of $\pi_{\Omega}$ in the case that the initial distance $R_0$ is fixed. The case where $R_0$ is averaged over the distribution $R_0^2 e^{-R_0^2/(2L^2)}/Z$ is treated exactly in the same way, but by keeping the average over $R_0$ at each step of the calculation. Then, we obtain the equation:
\begin{align}
&0=\int_0^{\infty}\frac{dt}{\psi^{5/2}}  \Bigg\{ \left(\frac{\mu_{i}^{\pi,0}R_{\pi}}{3}+ 
\frac{b_i\phi (\phi- e^{-\lambda_i t})}{\lambda_i } \right)e^{-\frac{R_{\pi}^2}{2\psi}} \nonumber\\
&- \int_a^{\infty}dR_0 \frac{R_0^2 e^{-\frac{R_0^2}{2L^2}}}{Z(a,L^2)}\frac{b_i \phi (\phi-e^{-\lambda_i t}) }{\lambda_i}  \left(1-\frac{R_0^2}{3\psi}
\right) e^{-\frac{\phi^2 R_0^2}{2\psi}}\Bigg\}.	\label{EquationFirstMomentDim3Averaged}
\end{align}
Taking into account the definition (\ref{DefinitionFunctionG}) of $G$, this equation is exactly (\ref{EquationFirstMomentDim3AveragedSimplified}) in the main text. 

\section{Asymptotic value of the moments $m_j^{\pi}$}
\label{AppendixAsymptoticValueMoments}
Here, we describe how to derive the asymptotic form (\ref{scalingRelationMomentsCyclization}) of the moments $m_j^{\pi}$, when $N\rightarrow\infty$ and $j\rightarrow\infty$ while  the value of $\tilde{a}=a/\sqrt{N}$ is held constant. We only study the case of end-to-end cyclization, with $p=1$ and $q=N$. It is straightforward to see that the equation (\ref{EquationFirstMomentDim3AveragedSimplified}) is invariant with $N$ if we use the rescaled variables $\tau=t/N^2$, $Y_{\pi}(\tau)=\sqrt{N}R_{\pi}(t)$, $\lambda_j=(j-1)^2\pi^2$, $b_j=-\sqrt{8/N}$ (if $j$ odd), $\Psi(\tau)=N\psi(t)$ and $\Phi(\tau)=\phi(t)$. The correct scaling for the moments is $M_j=m_{j+1}^{\pi}/N$ (we note that, trivially, $M_j=0$ for $j$ even).  With these rescaled variables, the equation (\ref{EquationFirstMomentDim3AveragedSimplified}) is independent on $N$. It can be developed in powers of $q$. The terms that do not contain $\exp(-j^2\tau)$ generate algebraic terms in $j$, and they must vanish, yielding a global condition on the unknown function $Y_{\pi}$. The remaining terms give the equation:
\begin{align}
&0= \int_0^{\infty}\frac{d\tau \ e^{-j^2\pi^2\tau}}{\Psi^{5/2}} \Bigg\{  \frac{M_j Y_{\pi}}{3} e^{-\frac{Y_{\pi}^2}{2\Psi}} + \frac{\sqrt{8}\Phi }{j^2\pi^2}\times\nonumber\\
&\Bigg[e^{-\frac{Y_{\pi}^2}{2\Psi}} \left(1-\frac{Y_{\pi}^2}{3\Psi}\right)
+ \left( \frac{Z(\tilde{a},\Psi)}{Z(\tilde{a},1)} -\frac{G(\tilde{a},\Psi)}{3\Psi Z(\tilde{a},1)}\right)\Bigg] \Bigg\}. 
\end{align}
All the terms in this equation can be evaluated in their short time limit. We introduce the simplifications: $\Psi(\tau)\simeq \kappa\tau$, $\Phi\simeq 1$, $Y_{\pi}\simeq \tilde{a}$, $Z(\tilde{a},\Psi)\simeq \tilde{a} \Psi e^{-\tilde{a}^2/(2\Psi)}$, and $G=(\tilde{a},\Psi)\simeq \tilde{a}^3 \Psi e^{-\tilde{a}^2/(2\Psi)}$. Keeping only the dominant terms for $\tau\rightarrow0$, we get:
\begin{align}
&0= \int_0^{\infty}\frac{d\tau}{\tau^{5/4}} e^{-j^2\pi^2\tau-\frac{\tilde{a}^2}{2\kappa\sqrt{\tau}}}\left(  \frac{M_j \tilde{a}}{3} - \frac{\sqrt{8} \tilde{a}^2}{3 jq^2\pi^2\kappa\sqrt{\tau} }\right)\label{Eq874} . 
\end{align}
Inverting this relation leads to:
\begin{align}
M_j= \left( \frac{\sqrt{8} \tilde{a}}{ j^2\pi^2\kappa } \right) \frac{\int_0^{\infty}d\tau \ \tau^{-7/4} \ e^{-j^2 H(\tau)}}{\int_0^{\infty} d\tau \ \tau^{-5/4} \ e^{-q^2 H(\tau) }  }\label{6873},
\end{align}
where we have posed $H(\tau)=\pi^2\tau+\tilde{a}^2/(j^2 2\kappa\sqrt{\tau})$. 
The integrals in (\ref{6873}) can be evaluated with the saddle point method.  The position of the saddle point is found by solving $\partial_{\tau}H(\tau^*)=0$, which gives $\tau^*= [\tilde{a}^2/(4\kappa j^2\pi^2)]^{2/3}$. Hence, we find that:\begin{align}
M_j= \frac{\sqrt{8} \ \tilde{a}}{ j^2\pi^2\kappa (\tau^*)^{1/2}}=\frac{2^{1/6} \tilde{a}^{1/3}}{\pi j^{4/3}}\label{78321},
\end{align}
where the last equality uses the value (\ref{ValueOfkappa}) of $\kappa$. Equation (\ref{78321}) is exactly the asymptotic form (\ref{scalingRelationMomentsCyclization})  in the main text. 

\section{Markovian expression of the effect of the distance between the reactive monomers}
\label{AppendixMarkovianEstimationDistanceReactiveMonomers}
\subsection{Case of two reactive monomers in the interior}
Here, we determine the asymptotic value of the reaction time when the reactive monomers are in positions $p$ and $q$ in the chain, with $\vert p-q\vert\ll N$. We work within the Markovian approximation. We introduce $s_p=p/N$ and $s_q=q/N$ the relative positions of the monomers in the chain, and $\Delta=\vert s_p-s_q\vert$, which represents the difference of curvilinear coordinate between the two reactive monomers, and $s^*=(p+q)/(2N)$, which is the average curvilinear coordinate of the two reactive monomers. In the limit $N\rightarrow\infty$, we can write $\phi(t)=\Phi(\tau)$, where $\tau$ is the rescaled time ($\tau=t/N^2$). The function $\Phi(\tau)$ is deduced from from the definition (\ref{DefinitionPhi}) and from the value (\ref{ExpressionCoefficient}) of the coefficients $b_i$ in the limit of large $N$:
\begin{align}
	\Phi(\tau)&=\frac{2N}{L^2}\sum_{j=1}^{\infty}\left[\cos(s_p \pi j)-\cos(s_q\pi j)\right]^2  \frac{e^{-j^2 \pi^2\tau}}{j^2\pi^2}.
\end{align}
Using the fact that $L^2=\Delta N$ and elementary trigonometry, we find:
\begin{align}
	\Phi(\tau)=	\frac{8}{\Delta}\sum_{j=1}^{\infty}\left[\sin\left(\frac{\Delta \pi j}{2}\right)\sin\left(\pi s^* j \right)\right]^2 \frac{e^{-j^2 \pi^2\tau}}{j^2\pi^2}\label{ExpressionOfPhi974}.
\end{align}
From Equation (\ref{MarkovianScalingLargeN}), we get that the reaction time (averaged over equilibrium initial conditions) in the limit of small target size ($a\ll \Delta\sqrt{N}$) scales as:
\begin{align}
	\frac{T}{N^2}= \int_0^{\infty}d\tau\left\{\frac{1}{[1-\Phi(\tau)^2]^{3/2}}-1\right\} \label{ExpressionOfT108}.
\end{align}
The right-hand side of this expression does not depend on $N$, but only on the parameters $\Delta$ and $s^*$, and we are looking for the limit $\Delta\rightarrow0$ at fixed $s^*$. Determining the limit of this expression (\ref{ExpressionOfT108}) in the limit $\Delta\rightarrow0$ requires to know the behavior of $\Phi$ in the same limit. Developing expression (\ref{ExpressionOfPhi974}) in the limit of small $\Delta$ at fixed $\tau$ leads to:
\begin{align}
	\Phi(\tau)\simeq  G(\tau,s^*)\Delta \ ; \ G(\tau,s^*)=2 \sum_{j=1}^{\infty} \sin(\pi s^* j)^2 e^{-j^2\pi^2 \tau}\label{SimplificationPhiTauLarge}.
\end{align}
Since by definition $\Phi(0)=1$,  expression (\ref{SimplificationPhiTauLarge}) cannot be valid for small values of $\tau$. It is in fact valid as long as $\tau\gg \Delta^2$. At smaller time scales, we write $\tau=u\Delta^2$, and we investigate the limit $\Delta\rightarrow0$ of  (\ref{ExpressionOfPhi974}) at fixed value of $u$. We find:
\begin{align}
	\Phi(u\Delta^2)\underset{\Delta\rightarrow0}{=}F(u)=4\int_0^{\infty}dx \sin\left(\frac{\pi x}{2}\right)^2\frac{ e^{-u x^2 \pi^2}}{x^2\pi^2}\label{SimplificationPhiTauSmall}.
\end{align}
The two expressions (\ref{SimplificationPhiTauLarge}),(\ref{SimplificationPhiTauSmall}) of $\Phi$ at different time scales can be matched by noting that, when $u\rightarrow\infty$ and $\tau\rightarrow0$, we have:
\begin{align}
F(u)\simeq\frac{1}{2\sqrt{\pi u}}= \frac{\Delta}{2\sqrt{\pi \tau }} \simeq \Delta\ G(\tau)\label{MatchingCondition}.
\end{align}
Let us introduce a parameter $\varepsilon$ that satisfies the condition $\Delta^2\ll \varepsilon\ll 1$ and is therefore in the intermediate time scale. Then, separating the integral (\ref{ExpressionOfT108}) into two pieces, and changing of variable in the first one, we get:
\begin{align}
	\frac{T}{N^2}\simeq  \Delta ^2 \int_0^{\frac{\varepsilon}{\Delta^2}}du&\left(  \frac{1}{[1-F(u)^2]^{3/2}}-1\right)\nonumber\\
	&+  \frac{3\Delta ^2 }{2}\int_{\varepsilon}^{\infty}d\tau [G(\tau,s^*)]^2. 
\end{align}
Because of the matching condition (\ref{MatchingCondition}), this expression does not depend on $\varepsilon$, and we obtain the following asymptotic expression for the reaction time:
\begin{align}
	T\simeq N^2 \Delta ^2 \left[ \frac{3}{4\pi} \ln \frac{1}{\Delta} + B(s^*) \right]\label{ExpressionMeanReactionTimeCloseReactiveMonomers},
\end{align}
where $B(s^*)$ is a numerical function of the average position of the monomers $s^*$ and is defined as $B=B_1+B_2$, with:
\begin{align}
B_1&=\lim_{\frac{\varepsilon}{\Delta^2}\rightarrow\infty} \left\{\int_0^{\frac{\varepsilon}{\Delta^2}} du \left[\frac{1}{[1-F(u)^2]^{3/2}}-1\right] - \frac{3}{8\pi} \ln \frac{\varepsilon}{\Delta^2} \right\},\nonumber\\
B_2&= \lim_{\varepsilon\rightarrow0}  \left\{ \frac{3}{8\pi} \ln \frac{1}{\varepsilon} + \int_{\varepsilon}^{\infty}d\tau \frac{3 [G(\tau,s^*)]^2 }{2} \right\}. 
\end{align}
The function $B(s^*)$ is represented in the main text on Fig. \ref{FigFonctionB}. We find numerically that, for $s^*\rightarrow0$, we have:
\begin{align}
	B(s^*)\simeq -\frac{3}{4\pi}\ln\frac{1}{2s^*}+\kappa_0, 
\end{align}
with $\kappa_0\simeq0.74$. Noting that, when $s^*\rightarrow0$, $s^*$ is approximately given by $\Delta/2$, we deduce that the reaction time for two close reactive monomers that are in the interior of the chain but close to the chain extremity is $T\simeq \kappa_0 (N\Delta)^2 $. 

\subsection{Case where one of the reactive monomers is at a chain extremity}
We now consider the case where the first of the two reactive monomers is located at one chain extremity. In this case, we have $s_p=0$, and the expression (\ref{ExpressionOfPhi974}) of $\Phi(\tau)$ becomes:
\begin{align}
	\Phi(\tau)=	\frac{8}{\Delta^2}\sum_{j=1}^{\infty}\left(\sin\frac{\Delta \pi j}{2}\right)^4 \frac{e^{-j^2 \pi^2\tau}}{j^2\pi^2}.
\end{align}
When $\Delta\rightarrow0$ at fixed $\tau$, we have: 
\begin{align}
	\Phi(\tau)\simeq \Delta^3 G_0(\tau) \ ; \ G_0(\tau)=	\frac{\pi^2}{2}\sum_{j=1}^{\infty}j^2e^{-j^2\pi^2\tau}\label{740}.
\end{align}
At the scale $\tau=u\Delta^2$, we get:
\begin{align}
\Phi(u\Delta^2)\underset{\Delta\rightarrow0}{=}F_0(u)=8\int_0^{\infty}dx\left(\sin\frac{x \pi }{2}\right)^4 \frac{ e^{-x^2\pi^2 u}}{x^2\pi^2}\label{7391}.
\end{align}
We note that $F_0(u)\sim 1/u^{3/2}$ when $u\rightarrow\infty$, while $G_0(\tau)\sim 1/\tau^{3/2}$ when $\tau\rightarrow0$. Then, using the approximations (\ref{740}),(\ref{7391}) of $\Phi$ at large and small time scales, we find that  in the limit $\Delta\rightarrow0$ the reaction time is given by:
\begin{align}
&	\frac{T}{N^2\Delta^2}\simeq  \nonumber\\
	&\int_0^{\frac{\varepsilon}{\Delta^2}}du\left( \frac{1}{[1-F_0(u)^2]^{3/2}}-1\right)
+  \frac{3\Delta^4}{2}\int_{\varepsilon}^{\infty}d\tau G_0(\tau)^2 \label{63742}.
\end{align}
We note that $F_0(u)\sim 1/u^{3/2}$ when $u\rightarrow\infty$, the that the first integral in (\ref{63742}) converges to a finite value in the limit $\varepsilon/\Delta^2\rightarrow\infty$. The second integral is divergent in the limit $\varepsilon\rightarrow0$ because $G_0(\tau)\sim 1/\tau^{3/2}$ for small $\tau$ ; however it remains of order $\Delta^4/\varepsilon^2$, which is small compared to one, and can therefore be neglected. Finally, we obtain:
\begin{align}
	T\simeq N^2\Delta^2 \int_0^{\infty}du \left\{ \frac{1}{[1-F_0(u)^2]^{3/2}}-1\right\}.
\end{align}
Numerically, we find $T=c N^2\Delta^2$, with $c\simeq0.38$. The fact that $c<\kappa_0$ indicates that the reaction between two reactive groups in the interior of the chain is slower than a reaction involving one monomer at the exterior of the chain. 


\end{document}